\documentclass{article}
\usepackage[a4paper,right=1in,left=1in,top=1in,bottom=1in]{geometry}

\usepackage{adjustbox}
\usepackage[T1]{fontenc}				                 
\usepackage{enumerate}	                                 
\usepackage{bm}								             
\usepackage{amsmath, amsfonts, amssymb}                  
\usepackage{mathrsfs}						             
\usepackage{graphicx,psfrag,epsf}			             
\usepackage{tabularx}						             
\usepackage{array}							             
\usepackage{float}							             
\usepackage{longtable}                                   
\usepackage[ruled]{algorithm2e}				             
\usepackage[subfigure]{tocloft} 		                 
\usepackage{makecell}						             
\usepackage[utf8]{inputenc}					             
\usepackage{multirow}						             
\usepackage[dvipsnames, svgnames, table]{xcolor}	     
\usepackage{units}							             
\usepackage{color, soul}					             
\usepackage[font=small,skip=0pt, margin=1cm]{caption}                
\usepackage{subcaption}                                  
\usepackage[final]{pdfpages}                            %
\usepackage[utf8]{inputenc}                              %
\usepackage[english]{babel}                              
\usepackage[square, numbers, comma]{natbib}                              
\usepackage{rotating}                                    
\usepackage{setspace}                                    
\usepackage{accents}                                     
\usepackage[title,titletoc,toc]{appendix}                
\usepackage{float}                                       
\usepackage{authblk}                                     
\usepackage[explicit]{titlesec}                          %
\usepackage[colorlinks=true, allcolors=Blue]{hyperref}   
\allowdisplaybreaks                                     %
\usepackage{arydshln}

\newcolumntype{P}[1]{>{\arraybackslash}m{#1}}
\definecolor{myrowcolor}{rgb}{.69,.7686,.8706}

\title{\Large
Moving Towards Automated Interstellar Boundary Explorer Data Selection with LOTUS}
\author[1,*]{\large Madeline A. Stricklin}
\author[1]{Lauren J. Beesley}
\author[1]{Brian P. Weaver}
\author[1]{Kelly R. Moran}
\author[1]{Dave Osthus}
\author[2]{Paul H. Janzen}
\author[3]{Grant David Meadors}
\author[4]{Daniel B. Reisenfeld}

\affil[1]{\footnotesize Statistical Sciences Group, Los Alamos National Laboratory, Los Alamos, NM, USA}
\affil[2]{Department of Physics and Astronomy, University of Montana, Missoula, MT, USA}
\affil[3]{Space Remote Sensing and Data Science Group, Los Alamos National Laboratory, Los Alamos, NM, USA}
\affil[4]{Space Science and Applications Group, Los Alamos National Laboratory, Los Alamos, NM, USA}
\affil[*]{Corresponding author: Madeline A. Stricklin, mstricklin@lanl.gov}
\date{}

\newcommand{\KitKat}[1]{\color{Gray}{\textbf{\begin{center} ********************************************* \\ 
*** BREAK POINT - KEEP GOING HERE *** \\ 
********************************************* 
\end{center}}}\normalcolor}

\begin{document}                                                         %
\maketitle

\abstract

The Interstellar Boundary Explorer (IBEX) satellite collects data on energetic neutral atoms (ENAs) that provide insight into the heliosphere, the region surrounding our solar system and separating it from interstellar space. IBEX collects information on these particles and on extraneous ``background'' particles. While IBEX records how and when the different particles are observed, it does not distinguish between heliospheric ENA particles and incidental background particles. To address this issue, all IBEX data has historically been manually labeled as ``good'' ENA data, or ``bad'' background data. This manual culling process is incredibly time-intensive and contingent on subjective, manually-induced decision thresholds. In this paper, we develop a three-stage automated culling process, called LOTUS, that uses random forests to expedite and standardize the labelling process. In Stage 1, LOTUS uses random forests to obtain probabilities of observing true ENA particles on a per-observation basis. In Stage 2, LOTUS aggregates these probabilities to obtain predictions within small windows of time. In Stage 3, LOTUS refines these predictions. We compare the labels generated by LOTUS to those manually generated by the subject matter expert. We use various metrics to demonstrate that LOTUS is a useful automated process for supplementing and standardizing the manual culling process. \\


\noindent \textit{Keywords:} Physics-Informed Machine Learning, Classification, Automation, Pattern Recognition, Random Forests, Heliophysics

\clearpage
\section{Introduction}                                                   %
The Interstellar Boundary Explorer (IBEX) satellite collects data on energetic neutral atoms (ENAs) that are believed to primarily originate from the boundary of the heliosphere \citep{Funsten:2009a, McComas:2009}. When a net electrically charged particle in the heliosphere exchanges charge with a net neutrally-charged particle in interstellar space, the resulting newly neutrally-charged particle is called an ENA. Because these ENAs no longer interact with magnetic fields, they travel in a ballistic trajectory from the point of neutralization. Depending on when and where a particle becomes neutralized, its ballistic trajectory may project it towards Earth. Some of these particles are detected by IBEX. 

Since 2008, IBEX has been collecting data on ENAs within certain energy ranges. These data are collected to inform our understanding of the boundary that exists between the heliosphere (i.e., our solar system) and interstellar space (i.e., what lies beyond our solar system) \citep{Funsten:2009a, McComas:2009}. In particular, these ENAs can be used to produce latitude/longitude maps of the ENA intensities (e.g., see \citep{Reisenfeld:2021, Osthus:2023}), which, in turn, allow us to study the global structure and dynamics of the heliosphere and its associated boundary (i.e., the heliopause) \citep{Funsten:2009b}.

IBEX collects information about ENAs as it orbits the Earth. The data it collects can be organized and understood based on where, when, and in what energy range the ENAs are observed. That is, the data it collects can be organized into \textit{orbit arcs}, \textit{angle bins}, \textit{times}, and \textit{electrostatic analyzer (ESA) steps}. An \textit{orbit arc} refers to an approximately 4.5 day period over which the spacecraft collects data from a fixed circular swath of the sky. IBEX observes the $360\circ$ swatch of sky that is roughly perpendicular to the Earth-Sun direction approximately every six months (i.e., views the entire sky over every six month period). During an orbit, as ENAs enter the device, the IBEX viewing \textit{angle} and \textit{time} of the detection are recorded, which provide information about how and when the ENA entered the detector. IBEX collects data across 60 different angle bins (6 degrees each), and numerous time intervals, which vary between orbit arcs. As ENAs enter the detector, they are captured by an ESA, which consists of two nested surfaces that set different energy passbands to limit the energy acceptance of the sensor to six different overlapping energy ranges (i.e., ESA steps 1-6). Each subsequent energy passband is broader than the previous, and more particles are detected in the higher energy passbands. Figure~\ref{SubFig:Raworbit} provides a visualization of what this data might look like for a given orbit and time period, across all angle bins and ESAs. Currently, only ESA steps 2 through 6 are used for end-data products, and so we only consider these ESA steps throughout the remainder of this paper.

However, not everything IBEX collects represents a true heliospheric ENA particle of interest. It also detects data on incidental background particles whose properties resemble those of ENAs. While the detector captures information about how many particles entered, and at what energy, time, and angle they were detected, it is not capable of cleanly distinguishing between true ENAs and background particles.  
 
Background interference can originate from many different sources. For example, a background signal may be the result of ENAs from the Earth's magnetosphere, ambient gas molecules that are ionized and accelerated within the sensor, ambient ions that are beyond the maximum rejection energy of the entrance subsystem, or coincidence events generated by penetrating radiation \citep{Funsten:2009b}. As it stands, these detectors are not able to distinguish the ENAs of interest from these background particles, and so a culling process is required to differentiate the observations consisting of counts of true ENAs (plus isotropic background), which are referred to as ``good times'' within the IBEX community, from the observations contaminated by anisotropic background particles, which are referred to as ``bad times'' within the IBEX community. Note that background particles are present in all data detected by IBEX. While some of these background particles exhibit predictable behavior and can easily be accounted for (we refer to these backgrounds as isotropic backgrounds), other background particles are less consistent in their behavior (we refer to these backgrounds as anisotropic backgrounds). Our goal is to separate out ENA particles (that exist with some quantity of isotropic background) from anisotropic background particles.

Currently, a manual culling process is used, in which a subject matter expert (SME) sorts through all the observations, and labels each as a ``good time'' or a ``bad time'' depending on whether or not it is believed that the resulting signal arose from true ENAs or from some background event. This manual culling process is extremely intensive and time-consuming, and the resulting ``good'' and ``bad'' determinations are inherently subjective. 

In this paper, we propose LOTUS\footnote{Leave Out The Ugly Stuff}, an automated culling process that serves to supplement the manual culling process, and propose a more standardized data culling approach. In Section~\ref{Sec:Data}, we discuss the different sources of data that are used to train LOTUS; in Section~\ref{Sec:Lotus} we describe the three stages that make up LOTUS; and in Section~\ref{Sec:ComparingLabs}, we compare the output of the manual culling process to the output of LOTUS and discuss the suitability of LOTUS as supplemental tool for the manual culling process.

\subsection{The Manual Culling Process and a Call    %
for Automation}                                      %
\label{SubSec:CallForAuto}                           %

The current process for distinguishing between background and ENA signals is subjective, time-intensive, and relies on manually-induced decision rules. At its simplest, the overall goal of the  culling process is to keep only the lowest count-rate intervals within an orbit. While the heliospheric signal remains relatively constant over the course of an orbit, background signal can be more sporadic in nature, often appearing as bursts in the data. To capture the heliospheric signal, the manual culling process involves removing signals associated with inconsistent events. Many background observations can be removed by following a set of expert induced decision rules. For example, the subject matter expert begins by removing observations associated with times and angles that are:
\begin{enumerate}[(1)]
    \item associated with obvious issues related to telemetry or instrument performance (e.g., ``spun times'' that occur when IBEX loses track of its point direction and does not correctly allocate particle detections into angles. These data can be ``despun'' or can be excluded from analysis); 
    \item detected when a celestial body is in the seven-degree field of view (FOV) of either IBEX sensor (e.g., when the Earth or Moon are captured in the FOV); 
    \item subjectively brighter than expected in multiple consecutive energy settings (e.g., when the counts of ENAs that are detected are higher than are expected).
\end{enumerate}
These three considerations are a good starting point for determining which observations correspond to true ENAs, and so are taken into account for all orbit arcs. However, this list is not exhaustive, and additional steps may need to be taken. For example, we may see instances where the Earth's magnetosphere is in the FOV, where the solar wind has interfered, or where observations between adjacent ESAs are inconsistent. These instances are not standard between all orbit arcs, and so must be considered on an orbit-by-orbit basis. 

Additional assumptions are also considered in culling the data. For example, if an anisotropic background signal is detected in ESA steps three and five, it should also be present in ESA step four. Additionally, the data are culled conservatively. In other words, the data are culled with the philosophy that it is better to label anything suspect as ``bad'', rather than potentially label any non-heliospheric signal as ``good''. Even though this conservative approach makes sense from a physical perspective, it does mean that potential ``good times'' may be omitted and that clear-cut classification rules may be more difficult to define. Finally, it is important to note that all data are manually culled twice, approximately a week apart, and the resulting labels are compared (Arriving at a final set of labels takes approximately an hour for each orbit arc. Considering that there are 954 individual orbit arcs in the considered dataset, that's approximately 954 hours spent culling data.). The ``better'' set is submitted as the finalized set of labels, where ``better" is subjectively determined by the same expert. As such, reproducing labels, even from a manual perspective, is difficult, and is never exactly equal between the two culling attempts (although the two attempts typically tend to be within some fraction of a standard deviation between one another). The presence of small additional backgrounds makes the manual culling process challenging to exactly replicate every time.

\begin{figure}
    \centering
    \begin{subfigure}[b]{0.90\textwidth}
        \centering
        \includegraphics[width=\textwidth]{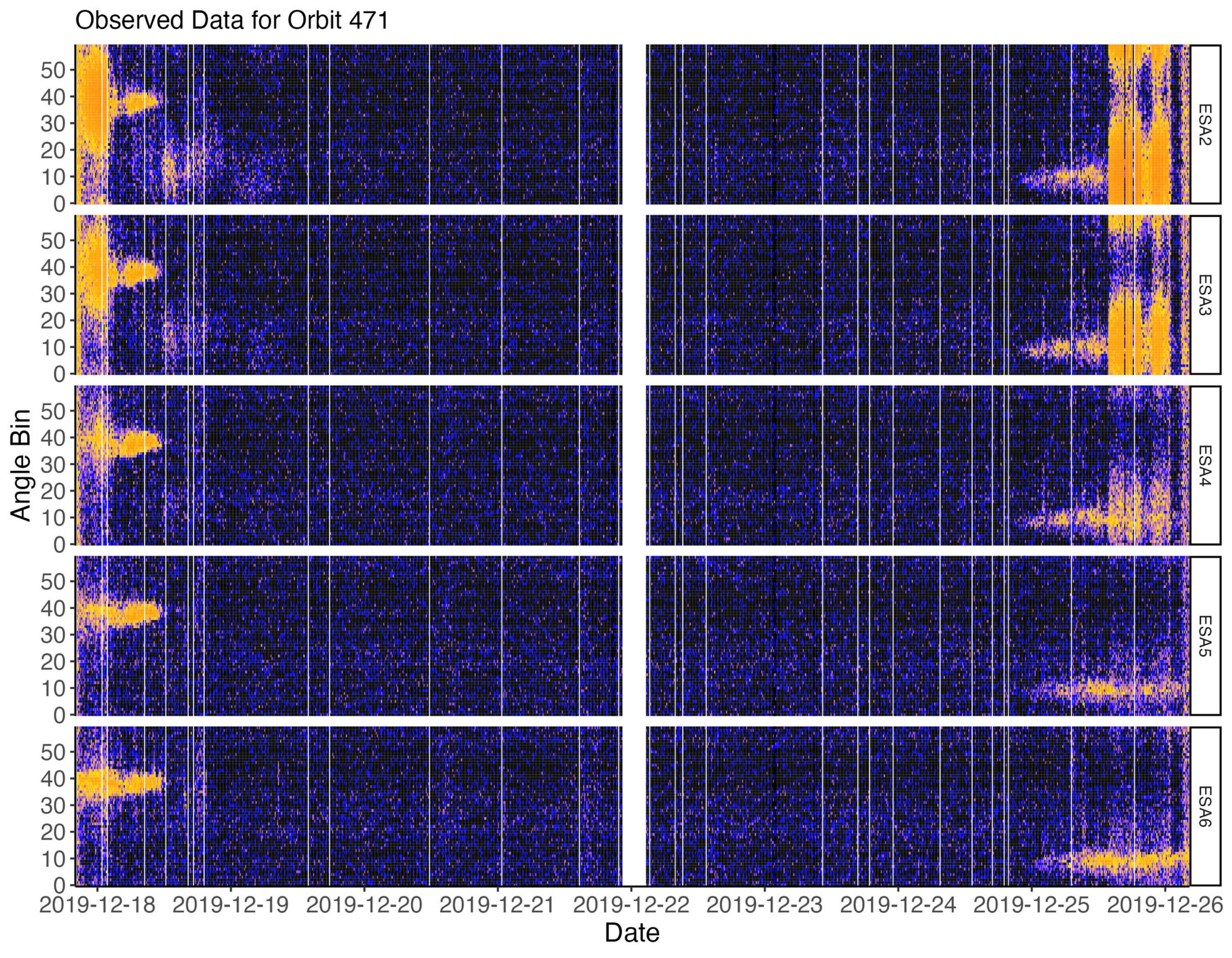}
        \caption{}
        \label{SubFig:Raworbit}
    \end{subfigure} \\
    \begin{subfigure}[b]{0.90\textwidth}
        \centering 
        \includegraphics[width=\textwidth]{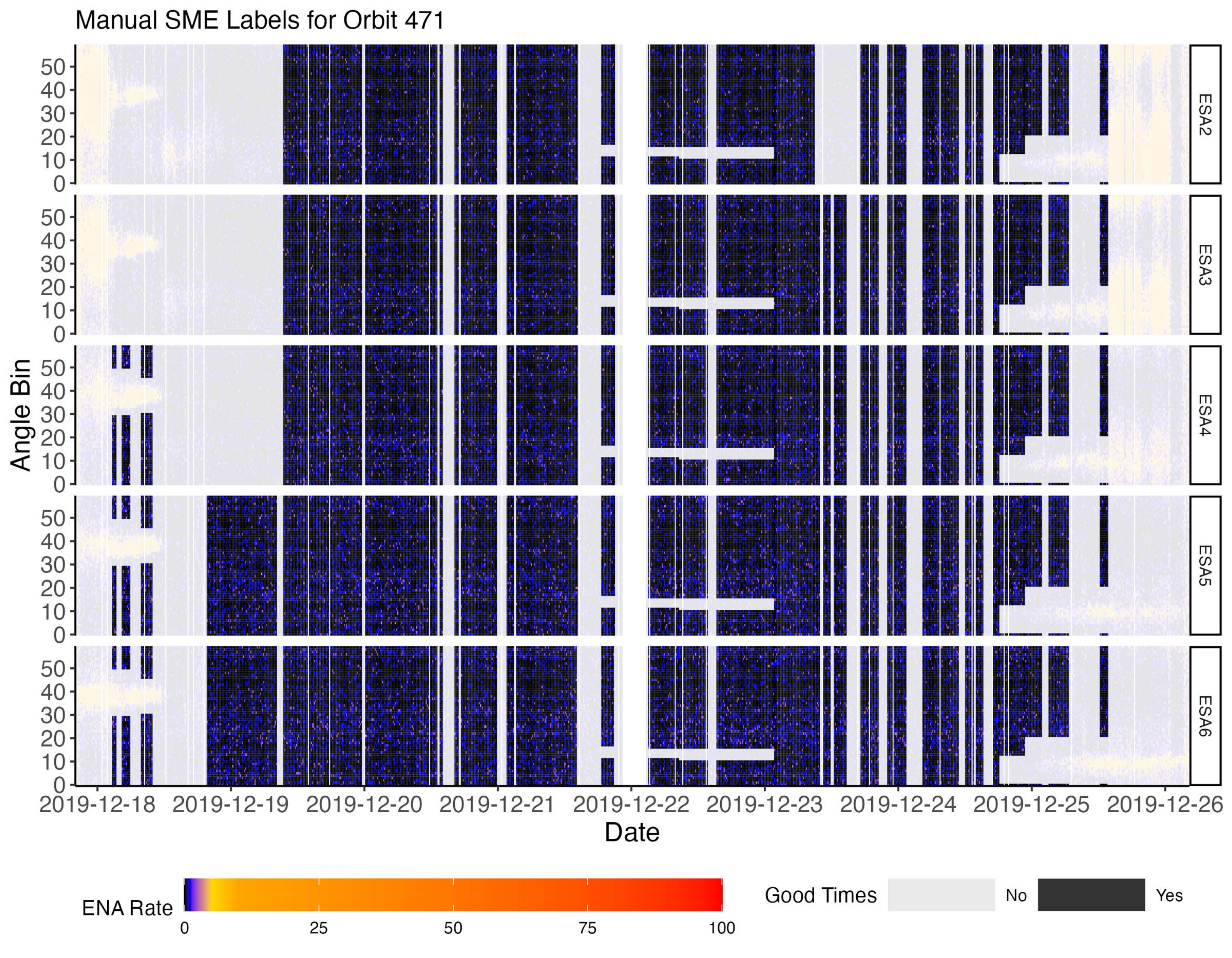}
        \caption{}
        \label{SubFig:Labeledorbit}
    \end{subfigure}
        \caption{ENA data collected for orbit 471, for ESA steps 2, 3, 4, 5, and 6. (a) Raw, unlabeled data collected by IBEX. (b) Manually culled data as determined by the SME. Pixel color corresponds to the count of ENA particles observed for a given time and angle bin. Greyed out areas in (b) correspond to ``bad times'', while maintained areas correspond to ``good times'', as identified by the SME.}
\end{figure}

Our goal is to propose an automated culling process that captures the considerations described above to create a faster, standardized approach for culling data while also maintaining the collective knowledge introduced by the SME in his labelling process. While this process can be greatly oversimplified by one rule (remove observations whose counts are ``too high'' based on some user-defined threshold), automating this process is not so straightforward, as human inconsistencies complicate this task. Even with over a decade of experience in labeling data, our SME acknowledges that there is no way to ensure he has been consistent in labelling data over the years, and that, even today, when he labels a single data set twice, there are inconsistencies in his own labels. Furthermore, the conservative manual culling process tends to result in entire sections of the data being labeled as ``good'' or ``bad'', rather than culling the data on a per-observation basis (see Figure~\ref{SubFig:Labeledorbit} for the block-like labelling structure used by the SME). This means that ``good'' observations may be thrown out if they fall too close to a set of obviously ``bad'' observations. Likewise, if a swath of ``bad'' observations is surrounded by ``good'' observations, the ``good'' observations may be foregone in order to ensure that ``bad'' observations are not included in the dataset. Unfortunately, this furthers the issue of inconsistency in labels between observations that may have similar features. This lack of repeatability within the manual culling process makes reproducibility all the more difficult, as any data we use to train an algorithm may be contradictory to other data in the set. As such, developing an automated culling method with neat discrimination rules is not a simple matter of feeding an algorithm a set of labeled training data. Instead, to replicate the manual process, additional considerations must be taken. While the manual culling process is time-consuming and subjective, we still aim to reproduce the results of our SME, who certainly is able to differentiate between useful ENA particles and anisotropic background particles. However, by automating the procedure, we can speed up the culling process as well as provide a standardized starting point for our SME to consider.

This paper presents LOTUS, an automated data processing alternative for determining whether or not a set of observations is made up of true ENAs or if it is a product of background interference. The automated alternative proposed in this paper is a supervised learning technique that can be used either on its own, or as a standardized starting point for the manual culling process. This automation will allow for a more timely, standardized culling process that results in less subjective label determinations. This method is useful for the current IBEX mission, and it also provides a basis for developing a method for classifying observations for the upcoming IMAP mission \citep{McComas:2018}. 

\section{Data Sources}                                                   %
\label{Sec:Data}                                                         %
The manual culling process relies on three data sources, described below, that we will also use for the automated culling process: (1) ENA detection data, (2) data from an external background monitor, and (3) information regarding the celestial bodies within the instrument's FOV. 

\subsection{ENA Detection Data}                      %
\label{SubSec:ENADat}                                %
IBEX collects data about the number of particles that may represent ENAs (with isotropic background) or anisotropic background data. This data is collected in approximately 4.5 day time periods, called \textit{orbit arcs}. Within each of these orbit arcs, IBEX captures data along a different circular slice of the 3-dimensional sky. These orbit arcs are further characterized by 60 six-degree \textit{angle bins} and many smaller \textit{time intervals}. At any given time, IBEX is collecting this data within one of six overlapping \textit{energy passbands} \citep{Funsten:2009b}, denoted ESA steps 1-6. Currently, a particle is counted towards the total number detected within an angle bin, time interval, and energy passband if it is detected by all three IBEX channel electron multiplier (CEM) sensors. In developing LOTUS, we consider data collected across 954 distinct orbit arcs consisting of 483,976 time intervals and 6 energy steps. The resulting ENA detection dataset contains over 174 million rows. 

\subsection{Background Monitor Data}                 %
\label{SubSec:BackgroundDat}                         %
In addition to collecting ENA detection data, the IBEX satellite includes a CEM background monitor that is sensitive to particles with energy above 14keV \citep{Allegrini:2009}. This monitor is unable to capture some of the sources of background present in the primary ENA detection data, and it is able to capture additional background interference that did not impact the primary ENA dataset. However, this background monitor data can still be viewed as an imperfect measure of the anisotropic background we want to screen out.

\subsection{SPICE Spatial Position Data}             %
\label{SubSec:SPICEDat}                              %
The third data source used in the manual culling process is information about whether the Moon or the Earth (or its magnetosphere) are visible in the instrument's FOV. While the Moon and the Earth are both sources for ENA emission, we are interested in studying the heliospheric ENAs generated much farther away. As a result, we discard observations that include these nearby ENA sources, because they may have unacceptably high background interference. The relative positioning of IBEX's FOV and these celestial bodies can be determined using SPICE, a set of computational tools developed by the US Navigation and Ancillary Information Facility (for more information, see \citep{Acton:1996, Acton:2018}). Supplemental Material~\ref{App:SpiceSpiceBaby} outlines the process used for determining which observations are made with the Moon or the Earth in view. 

While these data sources are extremely useful for beginning to develop an automated culling process, they are not sufficient on their own for transitioning from a manual culling process to an automated culling process (see Section~\ref{SubSec:CallForAuto}). In order to mirror the culling process used by our SME, it is helpful to define features that mirror the rules described in Section~\ref{SubSec:CallForAuto}. The three datasets considered above are useful for addressing (1) and (2) in Section~\ref{SubSec:CallForAuto} above. However, they do not provide sufficient information to fully address (3), or some of the other considerations mentioned in Section~\ref{SubSec:CallForAuto} (e.g., that observations should be correlated across consecutive ESA channels).

\subsection{Feature Engineering: Making the Most     %
            of IBEX's Data}                          %
\label{SubSec:FeatureEng}                            %
By using information captured in the three datasets listed above, we can engineer features that tell us about the spatial relationships that exist in the data and that better capture the SME's culling process. Such additional variables include ENA counts, means, and variances within time intervals, angle bins, and ESAs, as well as the counts of observations in neighboring observations (see Table~\ref{Tab:Features}). These types of features help tell us about spatial relationships that the SME takes into account and that might not otherwise be obvious in the available data. For example, by considering the information associated with a time interval and angle bin across all ESAs, we can connect these observations in a manner similar to that of our SME. That is, we can create a quantitative ``rule'' that allows us to label observations based on the information contained in neighboring ESAs. 

In the same way, by considering the observations directly adjacent to a given observation (that is, the counts associated with neighboring time intervals and angle bins), we can learn more about what a high or low particle count really indicates. For example, rather than eliminating an instance based solely off of the number of observed particles, we can determine if neighboring cells have homogeneously high counts (indicating that some sort of background interference is present), or if neighboring cells have lower counts (indicating that the high-count observation might just be a random fluctuation that isn't necessarily worth labelling as ``bad''). Table~\ref{Tab:Features} in Supplemental Material~\ref{App:RF_Feat} displays the 28 features considered in the final model with descriptions on how they are obtained. 

\section{LOTUS: An Automated Culling Process}
\label{Sec:Lotus}
LOTUS uses a random forest \citep{Hastie:2009} to generate predictions about whether a set of observations is ``good'' or ``bad'', based on the features displayed in Table~\ref{Tab:Features} in Supplementary Material~\ref{App:RF_Feat} (Recall that an observation is given by a 28-dimensional vector consisting of the features described in Table~\ref{Tab:Features}. This observation describes where, when, and how a particle count was observed, along with the label assigned by the SME, and additional spatial information. A single 28-dimensional observations corresponds to a single pixel in Figure~\ref{SubFig:Raworbit}.). Neural networks, support vector machines, and logistic regression \citep{Bishop:2006, Hastie:2009} were among other methods that were tested, but these methods demonstrated worse prediction capabilities than their random forest counterparts. Additionally, random forests provided the more nuanced added benefit that they most closely mirror the manual data culling process, which is ``tree-like'' in nature and is based on natural decision thresholds, as discussed in Section~\ref{SubSec:CallForAuto}. The final LOTUS model consists of three stages that allow for making predictions in a manner that reduces misclassification rates, while maintaining the classification structure induced by the SME (see Section~\ref{SubSec:CallForAuto} and Figure~\ref{SubFig:Labeledorbit}). We use orbit 471 as an example to demonstrate the different phases of LOTUS. 

\subsection{LOTUS Stage 1: Obtaining Probability of  %
Class with Random Forests on a Per-Observation Basis}%
\begin{figure}[H]
    \centering
    \begin{subfigure}[b]{0.75\textwidth}
        \centering
        \includegraphics[width=\textwidth]{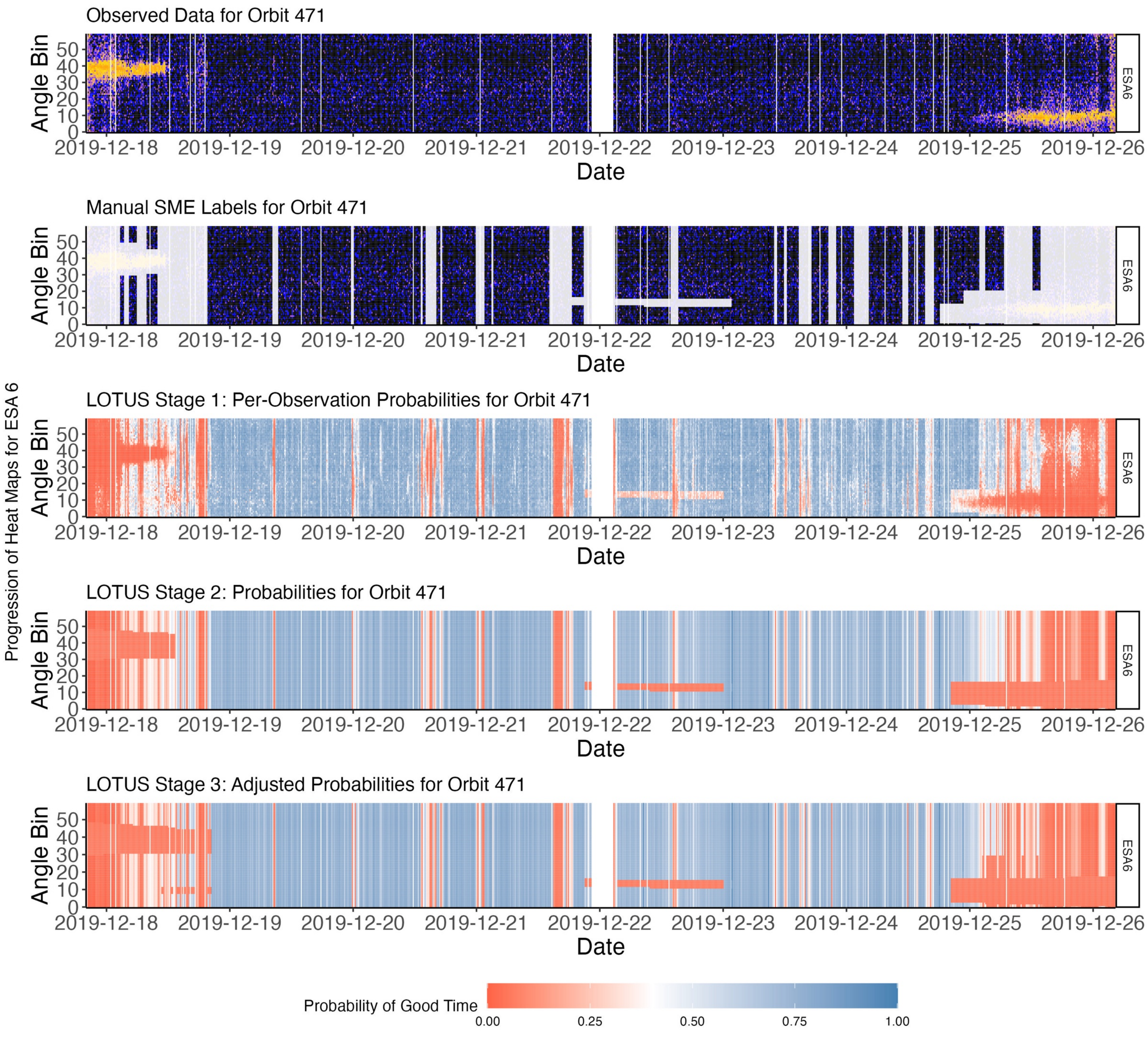}
        \caption{}
        \label{SubFig:471_6_Probs}
    \end{subfigure} \\
    \begin{subfigure}[b]{0.75\textwidth}
        \centering
        \includegraphics[width=\textwidth]{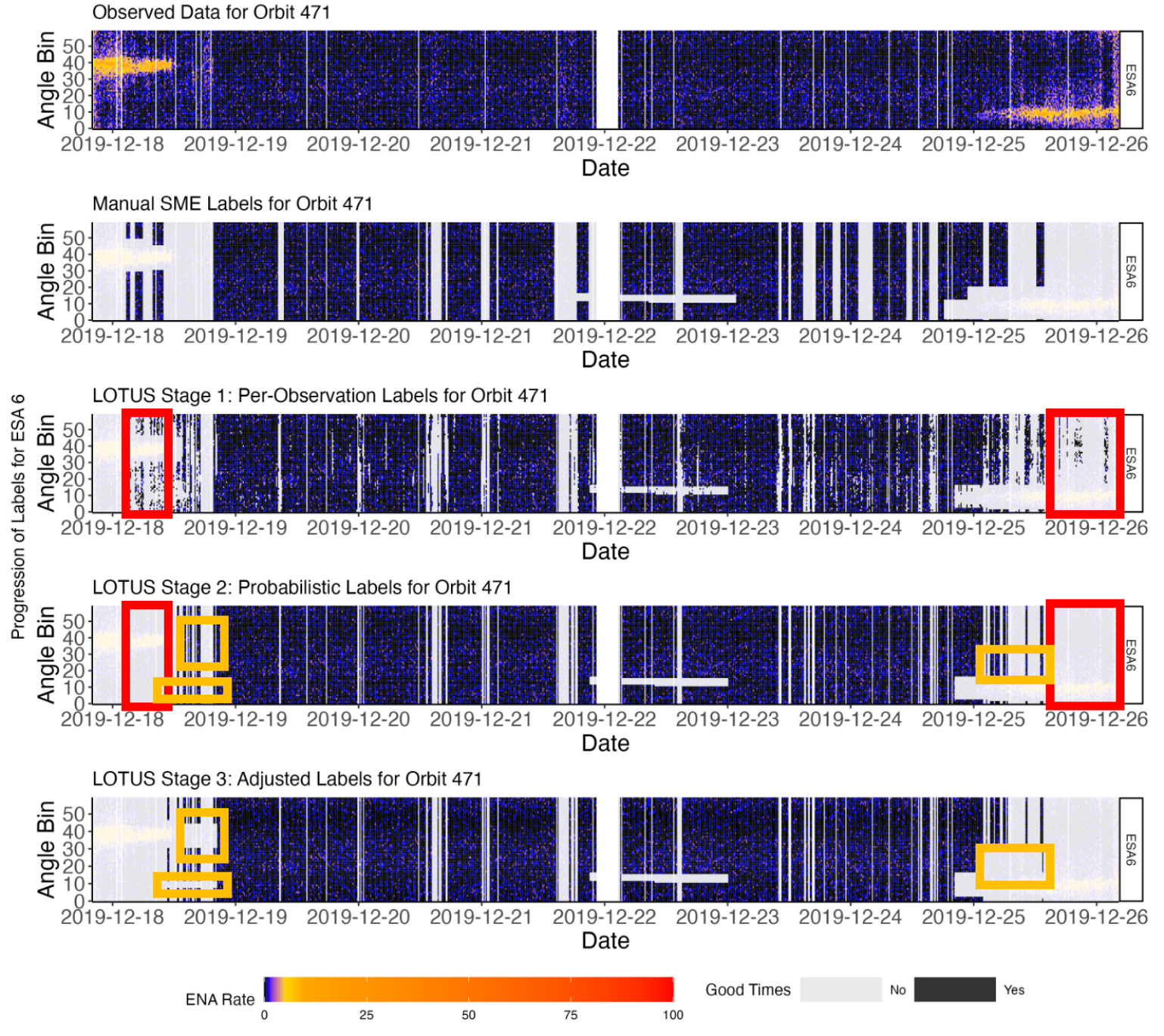}
        \caption{}
        \label{SubFig:471_6_Labs}
    \end{subfigure}
    \caption{Raw data, manual SME labels, and progression of LOTUS Stage 1, 2, and 3 for orbit 471, ESA step 6. Row 1 corresponds to the raw data observed by IBEX. Row 2 corresponds to the labels assigned by the SME. Rows 3, 4, and 5 correspond to the outputs of LOTUS Stage 1, 2, and 3, respectively. (a) Probabilistic progression of LOTUS. (b) Label progression of Lotus based on probabilistic output in (a). Red boxes correspond to changes that occurred between LOTUS Stages 1 and 2. Orange boxes correspond to changes that occurred between LOTUS Stages 2 and 3. Rows 1 and 2 are duplicated in both (a) and (b) for reader convenience. SPICE data is used to account for horizontal banding that may have been removed removed in LOTUS Stage 2.}
    \label{Fig:471_6_LOTUS}
\end{figure}

\label{SubSec:Lotus1}                                %
Stage 1 of LOTUS consists of fitting a random forest to predict whether an observation is a clean heliospheric observation (or is a ``good time'') or if it contains additional incidental backgrounds (or is a ``bad time'') as a function of the 28 features presented in Table~\ref{Tab:Features}. We use the \texttt{ranger} package in \texttt{R} \citep{R:2021, ranger:2017} to fit the LOTUS Stage 1 random forest. For a given orbit arc whose labels we wish to predict, we train a random forest on 250,000 randomly sampled observations. Several experiments were performed to determine a suitable number of randomly sampled observations. It was found that considering more than 250,000 randomly sampled observations did not add much value in terms of prediction capability, while considering less than 250,000 observations resulted in slightly reduced prediction capability. These observations are sampled to include observations from any other orbit arc than the one being considered, where an observation is defined by the 28-dimensional vector consisting of the features defined in Table~\ref{Tab:Features}. The output of this random forest is the probability than an observation is a true ENA particle, for each of the observations associated with the given orbit arc. As an example, consider orbit 471, consisting of orbit arcs 471a and 471b, together (see Figure~\ref{Fig:471_6_LOTUS}). In this example, the random forest outputs 300,971 predictions about whether the 300,971 observations are true ENA particles. This output can be seen in the third row of Figure~\ref{SubFig:471_6_Probs}. 

Row 3 of Figure~\ref{Fig:471_6_LOTUS} demonstrates the output of the random forest for orbit 471, ESA 6. In this figure, blue colors correspond to a higher probability that an observation is a true ENA particle, while salmon colors correspond to a lower probability. White areas correspond to the particles that are associated with the most uncertainty. Vertical white bars present in the raw data correspond to missing data areas in which IBEX was not collecting data. Similar plots for ESAs 2, 3, 4, and 5 can be seen in Supplemental Material~\ref{App:471_orbitPlots}.

We can see that the output of this random forest generally corresponds to how our SME might make predictions. That is, lower probability (or salmon) areas in \ref{SubFig:471_6_Probs} correspond to ``bad times'' (or grey) areas in \ref{SubFig:471_6_Labs}. We consider a cutoff of 0.40 as the threshold for whether or not an observation is a ``good time'' or a ``bad time'' (That is, observations whose probabilities are greater than or equal to 0.4 are labeled as ``good times'', while observations whose probabilities are less than 0.4 are labeled as ``bad times''. We choose 0.40 as the threshold, as this probability increased the percentage of observations that were labeled the same by both SME and LOTUS across all orbits compared to a ``standard'' threshold of 0.5). The resulting labels can be seen in row 3 of Figure~\ref{SubFig:471_6_Labs}. Using these labels, we can compare the output of the SME with that of LOTUS Stage 1. For example, we can see that LOTUS Stage 1 identifies the same count rate enhancement throughout the first half of orbit 471 that is identified by the SME. However, from a per-observation perspective, we also see that this area extends beyond that identified by the SME. Rather than exclude the entire area, as in the SME labeled data, LOTUS Stage 1 maintains the surrounding potentially ``good times''. 

\subsection{LOTUS Stage 2: Obtaining Class           %
Predictions on a Per-Time Basis}                     %
\label{SubSec:Lotus2}                                %
While the output of LOTUS Stage 1 is on a per-observation basis, we are more interested in observations made on a per-time interval basis. This allows us to more closely mimic the block-like structure that results from our SME's culling process. This structure is desirable from a physical perspective because it is difficult to justify a mechanism that would affect only one ESA, for only one angle bin and one time interval, as opposed to something slower and more gradual that would manifest over multiple ESAs, multiple angle bins, and multiple time intervals. We mimic this process by calculating the average probability that an observation is a ``good time'' within an entire time interval. By considering the probabilities over angle bins and within time intervals (i.e., by averaging over a column in Figure~\ref{Fig:471_6_LOTUS}), we begin to induce this block-like culling structure, as can be seen in Figure~\ref{SubFig:471_6_Probs} (see Figure~\ref{SubFig:Labeledorbit} for SME-labeled data product). 

Row 4 of Figure~\ref{Fig:471_6_LOTUS} demonstrates the aggregated probabilities and corresponding labels associated with ESA step 6 for orbit 471. While this process creates a product that resembles that of our SME, it is worth noting that aggregating the probabilities as such ends up eliminating lower-probability 
horizontal bands that occur. In order to ensure that these bands are maintained as a ``bad time'', we use our SPICE data to eliminate these areas; because these bursts are commonly associated with background interference due to the Earth or Moon, these areas can be easily eliminated. By considering the probability of a ``good time'' or ``bad time'' as before (where a ``good time'' has a probability of greater than or equal to 0.40), and by using the SPICE information to account for the horizontal banding that may have been removed when we aggregate across angle bins, we arrive at the labels displayed in row 4 of Figure~\ref{SubFig:471_6_Labs}. Areas designated by the red boxes in rows 3 and 4 of Figure~\ref{SubFig:471_6_Labs} indicate area that are removed as a function of the aggregation process used in LOTUS Stage 2.

\subsection{LOTUS Stage 3: Adjusting for Stage 1     %
Per-Observation Probabilities}                       %
\label{SubSec:Lotus3}                                %
In Stage 3, we make adjustments to account for consecutive low-probability areas detected in LOTUS Stage 1. This stage helps to detect additional regions that have very low probability of being a true ENA particle, but that may be eliminated or washed out when we average the probability across angle bin in LOTUS Stage 2. 

For example, we can see in Figure~\ref{SubFig:471_6_Probs} that LOTUS Stage 1 detects a region of ``bad times'' around time 2019-12-19 that is not removed in LOTUS Stage 2, due to there being high probability regions of ``good times'' (see the corresponding regions marked with the first two orange boxes in rows 4 and 5 of Figure~\ref{SubFig:471_6_Labs}). By identifying low probability regions that were not excluded and that exist over consecutive time bins, we can make an adjustment to remove these time bins. We can see a similar region is detected around 2019-12-25 (see the corresponding regions marked with the third orange box in rows 4 and 5 of Figure~\ref{SubFig:471_6_Labs}). By considering row 5 of Figure~\ref{SubFig:471_6_Probs}, we can see that some of the previously probable ``good times'' in LOTUS Stage 2 are now probable ``bad times'', bringing us closer into alignment with the assigned SME labels.

At this point, we can use these probabilities to obtain the final labels for the test orbit (orbit 471). As in LOTUS Stage 1 and 2, we use a probability of 0.4 as our threshold for whether or not a time bin is labeled as ``good'' or ``bad''. These final labels are presented in row 5 of Figure~\ref{SubFig:471_6_Labs}. Using this process, we can see that our final labels more closely resemble those of the SME, depicted in row 2 of Figure~\ref{SubFig:471_6_Labs}. 

A visual assessment of this orbit may indicate that we have missed the mark in some areas, potentially including too many good times. However, it is not the labels themselves that are the final target of analysis; in practice, observations are aggregated into estimates of ENAs observed per second for each angle bin (i.e., row). Different ``good time'' classifications identified by our SME and LOTUS could result in similar aggregations of ENAs per and similar performance in downstream analysis. To address these different analytical targets, we consider a variety of metrics for evaluating how well the output of LOTUS matches up with the output from our SME, both in terms of per-observation classification rates and in terms of similarity in aggregated ENA rate estimates. These diagnostics are explored in the following section.

\section{Comparing automated LOTUS labels to manual SME Labels}          %
\label{Sec:ComparingLabs}                                                %
In this section, we conduct a variety of analyses to determine whether or not LOTUS is a reasonable substitute for the manual culling process, both in terms of per-observation classification and in terms of downstream ENA rate estimates (i.e., row aggregations).  We can evaluate whether LOTUS' performance resembles that of the manual culling process using several different methods. In particular, we consider the following three methods: 
\begin{enumerate}[(1)]
    \item Correct Classification Rate;
    \item Ratios of ENA Rates (Manual vs. Automated); 
    \item Qualitative and Quantitative Comparisons of Downstream ENA Rate Maps.
\end{enumerate}
We compare each of these methods to a ``null model'' based on the intermediary point-by-point prediction step (i.e., LOTUS Stage 1). 

Together, these methods help us determine if the labels produced by  LOTUS are comparable to those of the manual culling process. For example, misclassification rates alone do not tell a complete story in terms of the quality of our data classifications; rather, we are more directly interested in the subsequent ENA rate estimates used to estimate global heliospheric latitude/longitude maps of ENA rate/intensity. While the per-observation misclassification rate can provide initial insight into the usefulness of a given model, it is possible to generate ENA rate maps from an automated culling process that resemble maps from a manual culling process, even if these two sets of maps are derived from different sets of ``good time'' or ``bad time'' labels. 

\subsection{Correct Classification Rate}             %
\label{SubSec:MisclassRate}                          %
Initially, it might make sense to consider the percent of observations that are classified as either ``good'' or ``bad'' by both the manual and automated culling processes (i.e., the correct classification rate). When only this method is considered, the automated culling process appears to slightly miss the mark. For example, the average correct classification rate for orbit 471 is 86.91\% (see Table~\ref{Tab:ClassificationRates} for a breakdown of the correct classification rate for orbit 471 by ESA step). This closely corresponds to 83.90\%, the average correct classification rate across all orbit arcs (see Table~\ref{Tab:ClassificationRates} for a breakdown of the correct classification rate across all orbit arcs by ESA steps). When we consider all orbit arcs, it it also helpful to consider the 95\% highest density interval of the correct classification rates. Over the total 928 unique orbit arcs, the 95\% highest density interval ranges from 70\% to 100\%. We can further break down these classification rates into sensitivities and specificities, which give the ``true positive'' and ``true negative'' rates, respectively. That is, the sensitivity gives the probability of a LOTUS designated ``good time'', given an SME designated ``good time'', while the specificity gives the probability of a LOTUS designated ``bad time'', given an SME designated ``bad time''. These rates are provided in Table~\ref{Tab:ClassificationRates}, and indicate that, in general, LOTUS Stage 3 ``bad times'' truly correspond to the SME labeled ``bad times'', and that LOTUS Stage 3 may be missing some of the ``good times'' that are picked up by the SME.

\begin{table}[H]
    \centering
    \begin{tabular}{|c|c|c|c|c|c|c|}
        \hline
        \rowcolor{myrowcolor!30} & Statistic & ESA Step 2 & ESA Step 3 & ESA Step 4 & ESA Step 5 & ESA Step 6  \\
        \hline \hline 
        \parbox[t]{2mm}{\multirow{3}{*}{\rotatebox[origin=c]{90}{471}}} & Accuracy & 86.0 & 87.5 & 82.9 & 89.1 & 89.1 \\
        & Sensitivity & 82.3 & 85.5 & 79.2 & 88.2 & 88.6 \\
        & Specificity & 91.6 & 90.6 & 91.1 & 91.1 & 90.3 \\
        \hline
        \parbox[t]{2mm}{\multirow{3}{*}{\rotatebox[origin=c]{90}{All}}} & Accuracy & 82.9 & 84.5 & 83.8 & 84.2 & 83.1 \\
        & Sensitivity & 64.4 & 74.0 & 75.9 & 78.0 & 75.1 \\
        & Specificity & 92.4 & 90.5 & 90.3 & 89.5 & 89.7 \\
        \hline
    \end{tabular}
    \caption{Percentage of observations whose predicted LOTUS Stage 3 labels match the manually assigned SME labels, broken down by sensitivity and specificity, for orbit 471 (first three rows) and all orbit arcs (last three rows), by ESA Step.}
    \label{Tab:ClassificationRates}
\end{table}



\subsection{ENA Rates}                               %
\label{SubSec:ENARates}                              %
Success of our task does not depend solely on the ability of an algorithm to correctly label an observation as ``good'' or  ``bad''.  Ultimately, the observations that have been deemed ``good'' go on to be used for map-making \citep{Osthus:2023, Reisenfeld:2021} in the form of aggregates called ENA rates. So, while an observations labelled as ``good'' or ``bad'' by an algorithm may not correspond exactly to the  ``good'' or ``bad'' label assigned by our SME, the resulting ENA rates may still be similar, and a sufficient map may still be produced. As a result, misclassification rates are insufficient on their own to evaluate an automated process's performance. 

Aggregates of the ENA counts across time and within an angle bin are a more reliable gauge for whether or not the automated process is replicating the manual process. These aggregates, or ENA rates, are calculated for each angle bin and consider the difference between the ratio of the number of direct events (i.e., quantified ``good times'') to the exposure time (i.e., length of time accounted for by ``good times'') and the isotropic background rate (which is estimated using the ``good times'' and other coincidence data products from the considered orbit). The number of direct events corresponds to the number of observed particles that have been labeled as ``good times'', the exposure time corresponds to the length of time associated with the identified ``good times'' in seconds, and the background rate is the number of isotropic background particles per second. We can represent this ENA rate (for a given angle bin) by
\begin{eqnarray*}
    \text{ENA Rate} = \frac{\text{Total ``Good Time'' Counts}}{\text{Total ``Good Times''}} - \text{Isotropic Background}.
\end{eqnarray*}

\begin{figure}[H]
    \centering
    \includegraphics[width=\textwidth]{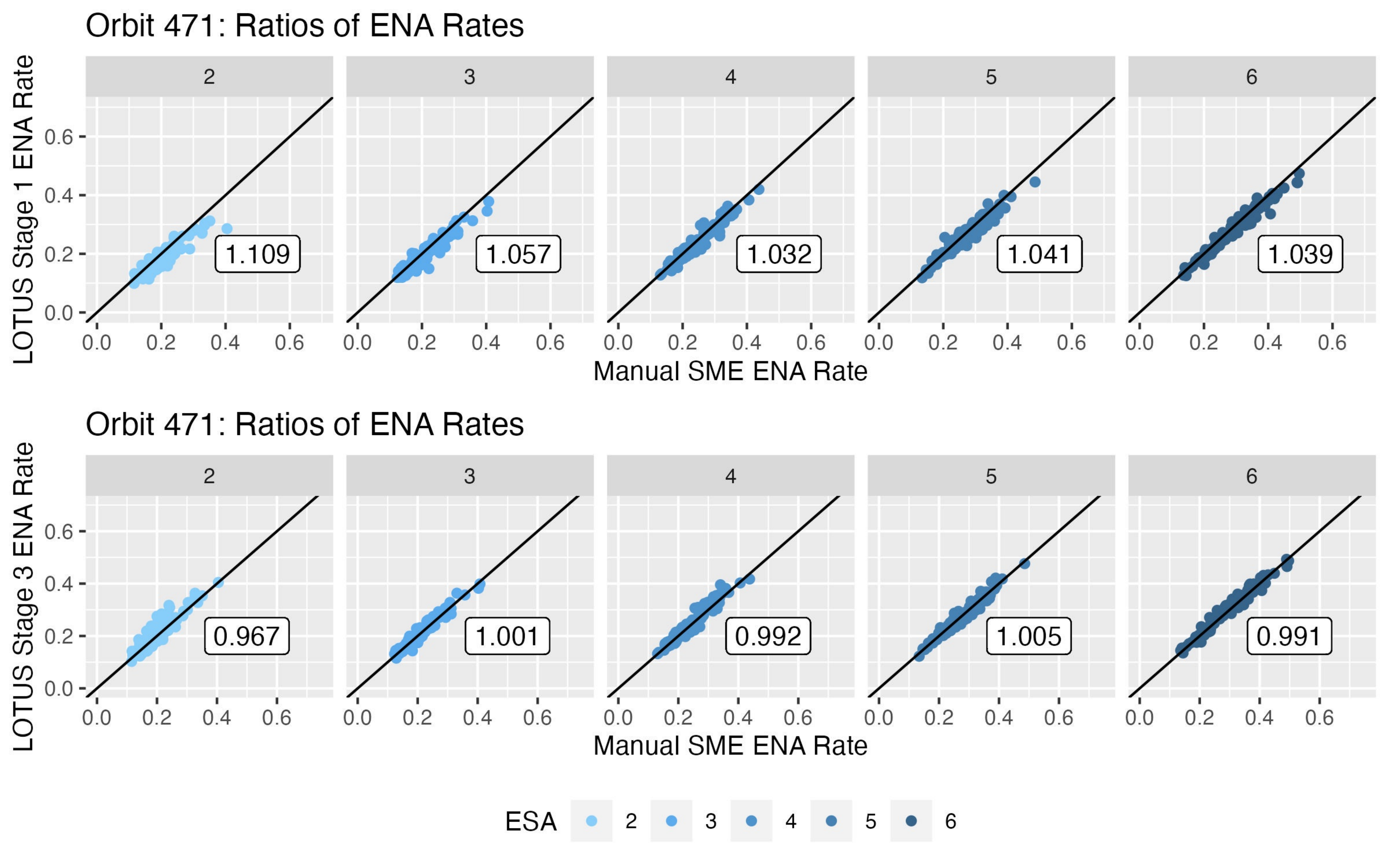}
    \caption{Manual SME ENA rate versus LOTUS ENA rate for orbit 471. Each point corresponds to an SME ENA rate on the x-axis and a LOTUS ENA rate on the y-axis, for a given angle bin. Columns correspond to ESA steps 2, 3, 4, 5, and 6, respectively. Labels correspond to the average ENA rate ratio for the given ESA step and given LOTUS stage.}
    \label{Fig:471_ENARatios}
\end{figure} 

Figure~\ref{Fig:471_ENARatios} shows the ENA rates that result from the manually assigned SME labels versus the ENA rates that result from each stage of LOTUS. The columns of Figure~\ref{Fig:471_ENARatios} correspond to ESA steps 2 through 6, while the rows correspond to LOTUS Stages 1 and 3. Each plot also provides the average ratio of the manual SME ENA rate to the LOTUS ENA rate. From this figure, we can see that the ratio becomes closer to 1.00 as we move from LOTUS Stage 1 to LOTUS Stage 3, and that the corresponding ratios become more clustered about the center of the line of equality (the line given by $x=y$, which corresponds to a perfect, one-to-one association between the manually culled SME ENA rate and the automatically culled LOTUS ENA rate). Additionally, we see that LOTUS Stage 1 tends to underestimate the ENA rate. LOTUS Stage 3 corrects for this underestimation. 

\begin{figure}[H]
    \centering
    \includegraphics[width=\textwidth]{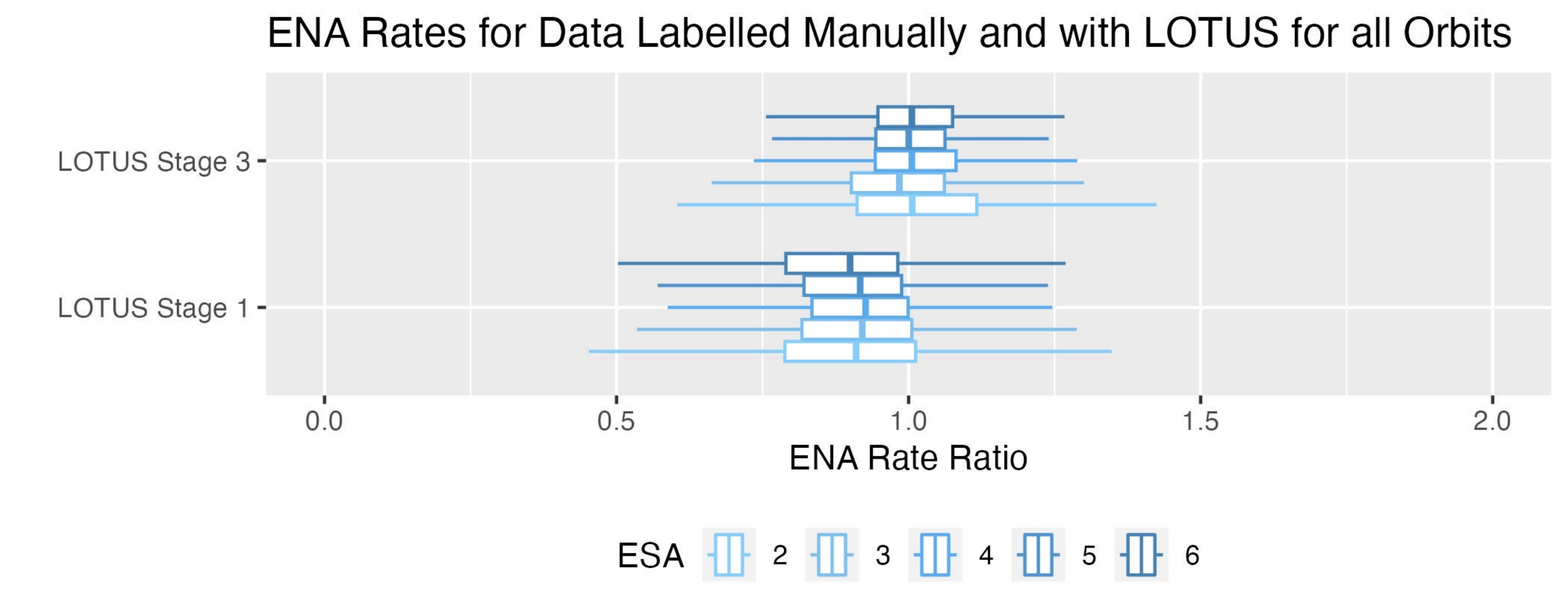}
    \caption{Boxplots of LOTUS/SME ENA rates for LOTUS Stages 1, 2, and 3, broken down by ESA step.}
    \label{Fig:All_ENARatios}
\end{figure}

We can also consider how the LOTUS ENA rates compare to those resulting from our SME across all orbits. Figure~\ref{Fig:All_ENARatios} depicts the box plots of the ratios of the ENA rates that result from the manual SME labels to the ENA rates that result from LOTUS Stages 1 and 3. As in the case for orbit 471, we can see that as we move from LOTUS Stage 1 to LOTUS Stage 3, the box plots of ENA rates begin to converge to a ratio of 1.00, indicating that the performance of the model is improving from one stage of LOTUS to the next.

\begin{figure}[H]
    \centering
    \includegraphics[width=0.92\textwidth]{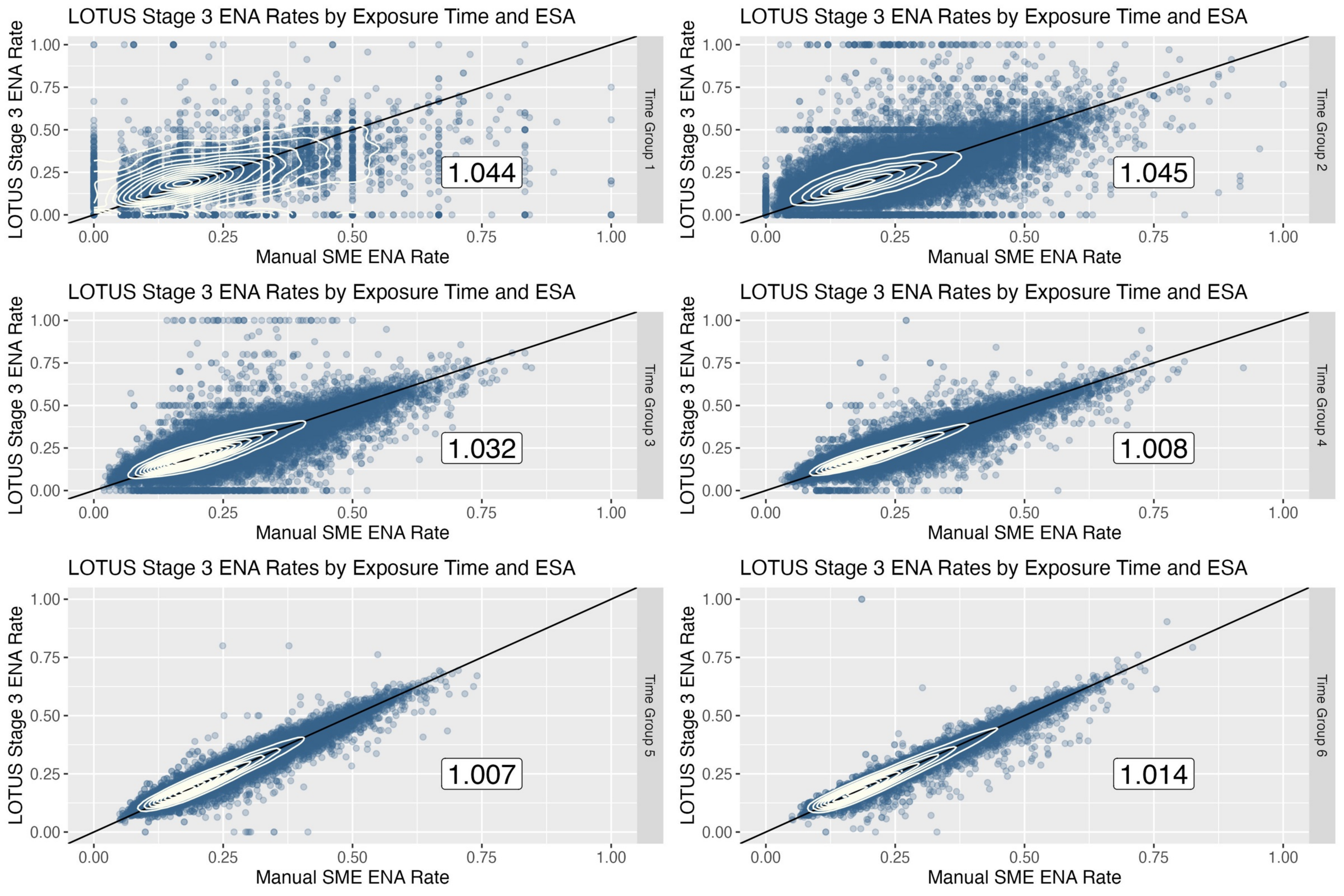}
    \caption{Manual SME ENA Rate compared to LOTUS Stage 3 ENA Rate. Each facet corresponds to one of the time groups defined in Table~\ref{Tab:ExposureTime}. Contours show distribution of clouds of points obtained from a 2D kernel density estimation \citep{Venabls:2013}.}
    \label{Fig:All_ENARatios_Time}
\end{figure}

Additionally, we consider the scatter plots of the Manual SME ENA rates versus the LOTUS Stage 3 ENA rates in Figure~\ref{Fig:All_ENARatios_Time}. In this particular plot, we consider the breakdown of points by ESA, as before, but we also break down the points by time groups. Because observations with longer exposure times contribute more heavily towards the final sky maps, while observations with shorter exposure times contribute less heavily towards the final sky maps, we investigate the breakdown of the ratios of the manual SME ENA rates to the LOTUS Stage 3 ENA rates. More specifically, we define time groups associated with the 10\%, 25\%, 50\%, 75\%, and 90\% quantiles of exposure time (see Table~\ref{Tab:ExposureTime} for the exposure times corresponding to each quantile). These results are presented in Figure~\ref{Fig:All_ENARatios_Time}.

\begin{table}[H]
    \centering
    \begin{tabular}{|c|c|c|c|c|c|c|}
    \hline
        \rowcolor{myrowcolor!30} Time Group & 1 & 2 & 3 & 4 & 5 & 6 \\
        \hline
         \rowcolor{myrowcolor!30} Quantile & 10\% & 25\% & 50\% & 75\% & 90\% & 100\%  \\
         \hline \hline 
         Exposure Time (secs) &  $<21$ & $21 - 82$ & $83-167$ & $168-276$ & $277 - 373$ & $>373$ \\
         \hline
    \end{tabular}
    \caption{Exposure time groups (in seconds) for 10\%, 25\%, 50\%, 75\%, 90\%, and 100\% quantiles.}
    \label{Tab:ExposureTime}
\end{table}

Figure~\ref{Fig:All_ENARatios_Time} demonstrates that the clouds of points become more tightly clustered as we progress through the different exposure time groups, as evidenced by the density contour lines overlaid in the clouds of points. This indicates that the output of LOTUS Stage 3 is more likely to match the output of the SME for the observations that are most important in the map making process. 

\subsection{Sky Maps}                                %
\label{SubSec:SkyMaps}                               %
Finally, we consider the sky maps themselves to determine whether LOTUS is a reasonable surrogate for the manual culling process implemented by our SME. IBEX sky maps allow us to visualize the ENA rates observed by the IBEX satellite in all directions \citep{Funsten:2009b}. These maps are typically pixelated in nature, due to being generated at 6-degree resolution.

\begin{figure}[H]
    \centering
    \includegraphics[scale=0.15]{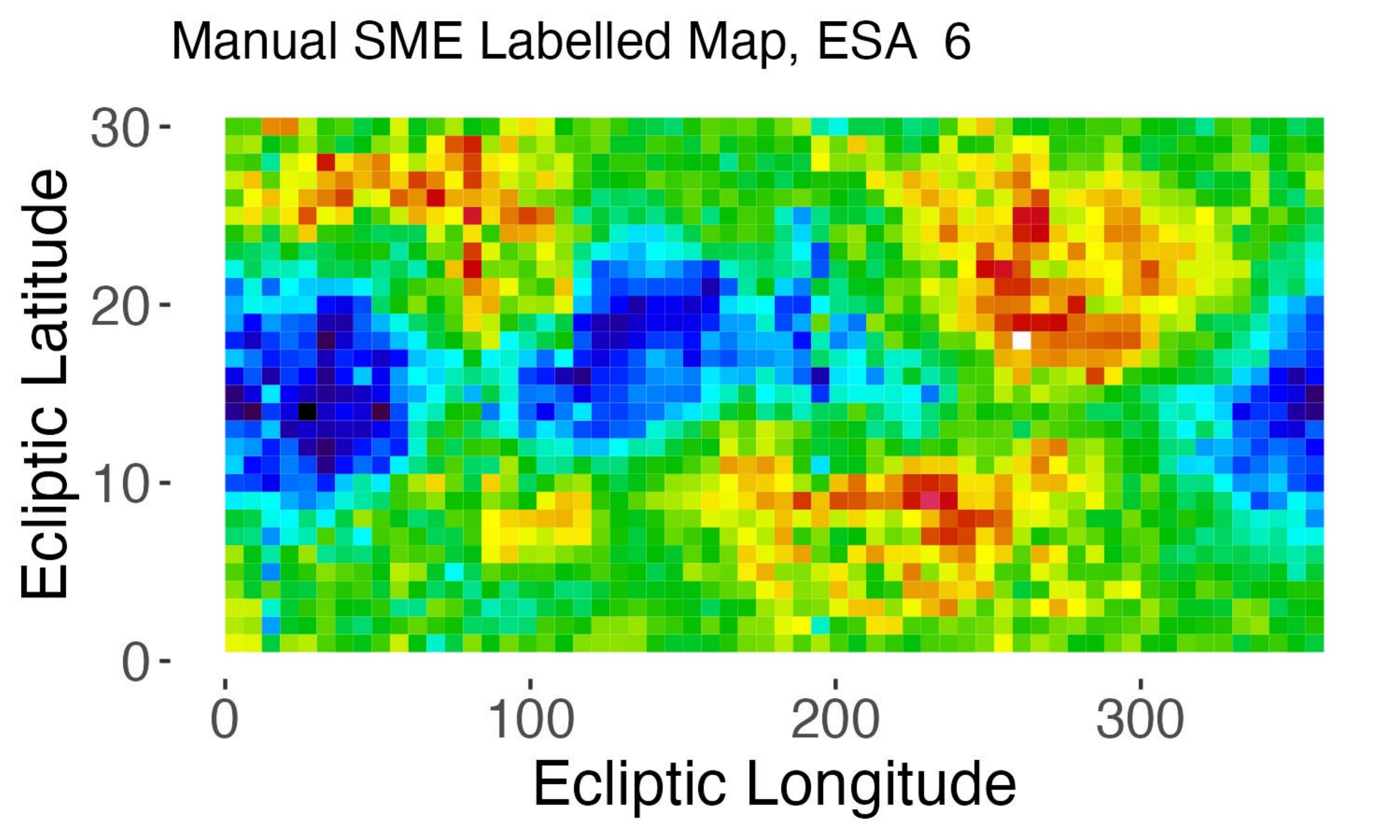} \\
    \includegraphics[scale=0.12]{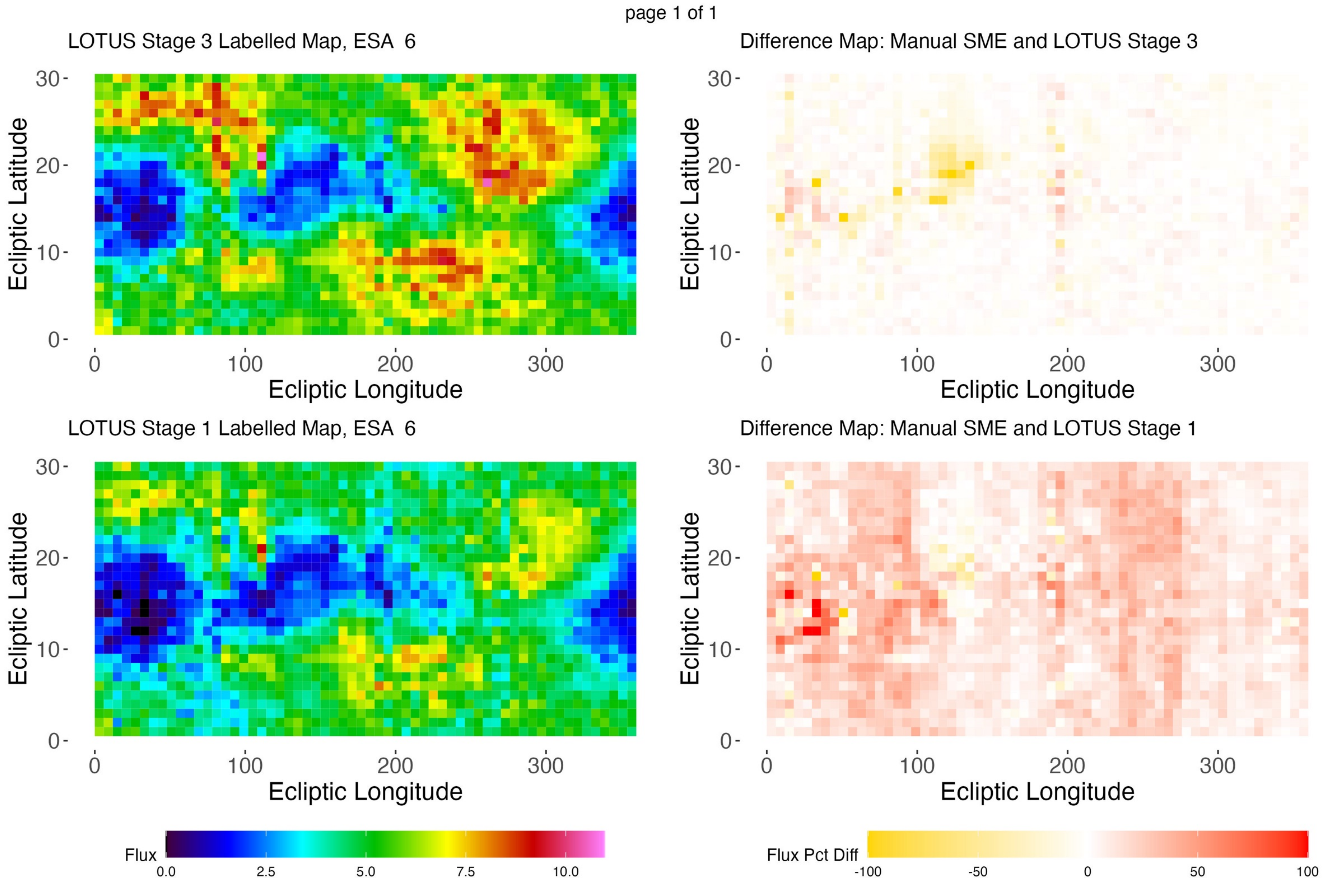}
    \caption{ISOC-like sky map for 2019B, ESA step 6. Rows correspond to maps resulting from manually culled labels, LOTUS Stage 3 labels, and LOTUS Stage 1 labels, respectively. Column 1 corresponds to the maps generated by the ENA rates for each model and column 2 corresponds to the percent difference between the manual SME Labelled map and LOTUS Stages 1 and 3 maps.}
    \label{Fig:2019B_ISOC_6}
\end{figure}

\begin{figure}[H]
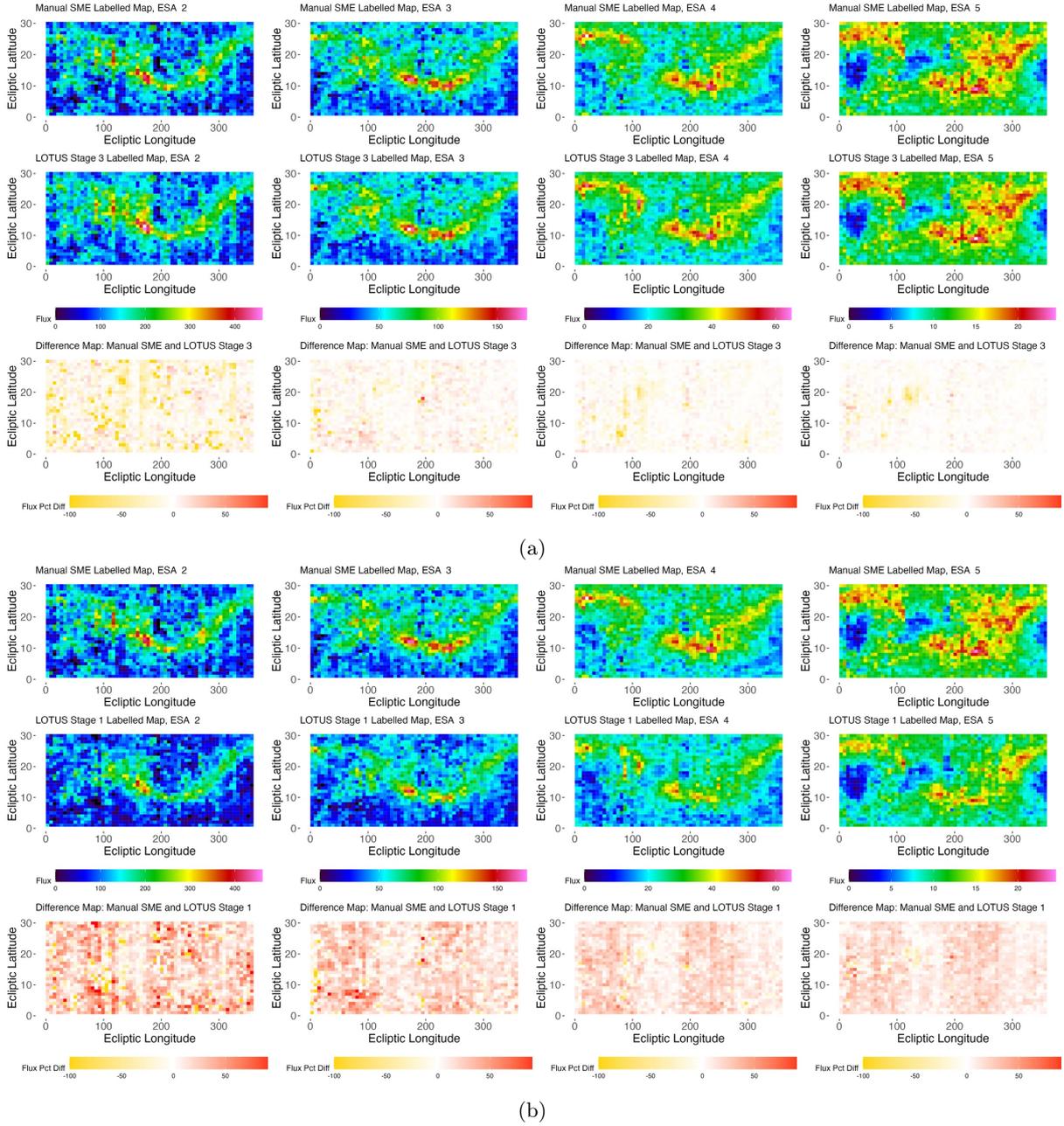

    \centering
    \begin{subfigure}[b]{\textwidth}
        \centering
        \includegraphics[width=\textwidth]{images/Fig7a_ISOC_Maps_L3_2019B.pdf}
        \caption{}
        \label{Fig:2019B_ISOC_other_a}
    \end{subfigure} \\
    \begin{subfigure}[b]{\textwidth}
        \centering
        \includegraphics[width=\textwidth]{images/Fig7b_ISOC_Maps_L1_2019B.pdf}
        \caption{}
        \label{Fig:2019B_ISOC_other_b}
    \end{subfigure}
    \caption{ISOC-like sky maps for 2019B, ESA steps 2, 3, 4, and 5 for (a) LOTUS Stage 3 and (b) LOTUS Stage 1. Rows correspond to maps resulting from manually culled labels, LOTUS labels, and percent difference maps, respectively.}
\end{figure}

We begin by considering a map making method that is similar to the one that is operationally used today by the IBEX Science Operation Center (ISOC) (note that these maps consider the uncorrected flux, rather than the ENA rate, which is a function of both ESA step and ENA rate). While more than 25 different maps exist for each ESA step when we consider all of the data, we will consider ISOC map 2019B (i.e., data collected in the second half of 2019) as an example for the remainder of this section. We choose this map since the orbit arc considered in the preceding sections of this paper (i.e., orbit 471) contributes to this map (Note that multiple orbit arcs contribute to each map. While orbit 471 is a contributor to map 2019B, it is one of 20 separate orbits that are used to construct map 2019B.). Figure~\ref{Fig:2019B_ISOC_6} displays the maps that result from the manual labelling process, from LOTUS Stage 3, and from LOTUS Stage 1 for ESA step 6. While we are more interested in the comparison between the maps that result from the manual labels and from LOTUS Stage 3 (the ultimate result of LOTUS), we also include the map that results from the output of LOTUS Stage 1. This helps to emphasize that a per-observation labelling scheme is not as effective as one that considers labels across angle bins, and within time bins. Similar maps for ESA steps 2 through 5 can be found in Figures~\ref{Fig:2019B_ISOC_other_a} and~\ref{Fig:2019B_ISOC_other_b}. 

A visual assessment of Figures~\ref{Fig:2019B_ISOC_6}, \ref{Fig:2019B_ISOC_other_a}, and \ref{Fig:2019B_ISOC_other_b} indicates that LOTUS does a satisfactory job at replicating the map produced from the manually assigned labels. While we see some areas of discrepancy (see areas of maps associated more saturated values of the differences map), we see that the overall shapes of the different structures in the LOTUS Stage 3 map are approximately the correct shape and size, with approximately the same intensities as those associated with the manual labels.

\begin{figure}[H]
    \centering
    \includegraphics[scale=0.15]{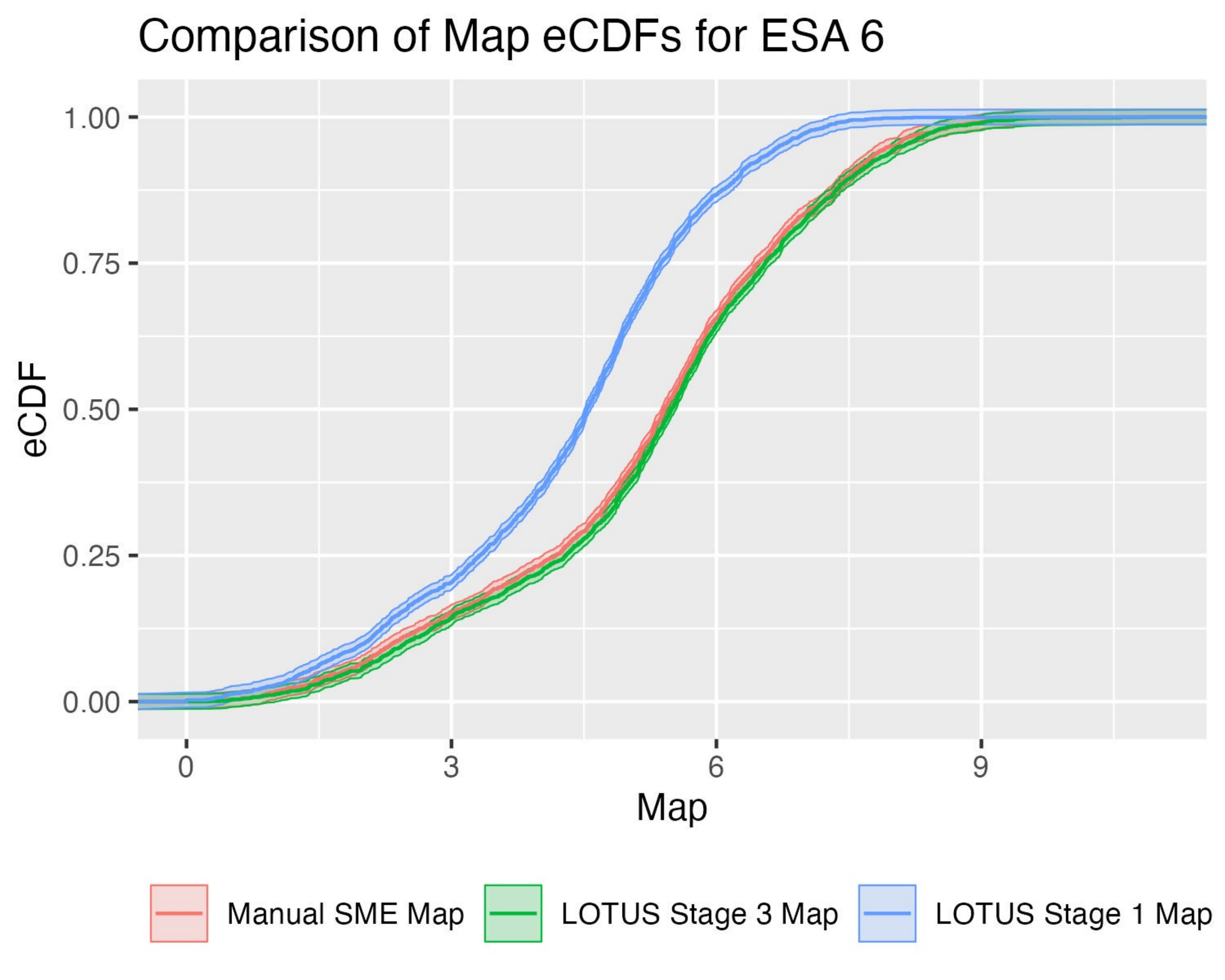} \nonumber \\
    \vspace{3mm}
    \includegraphics[width=\textwidth]{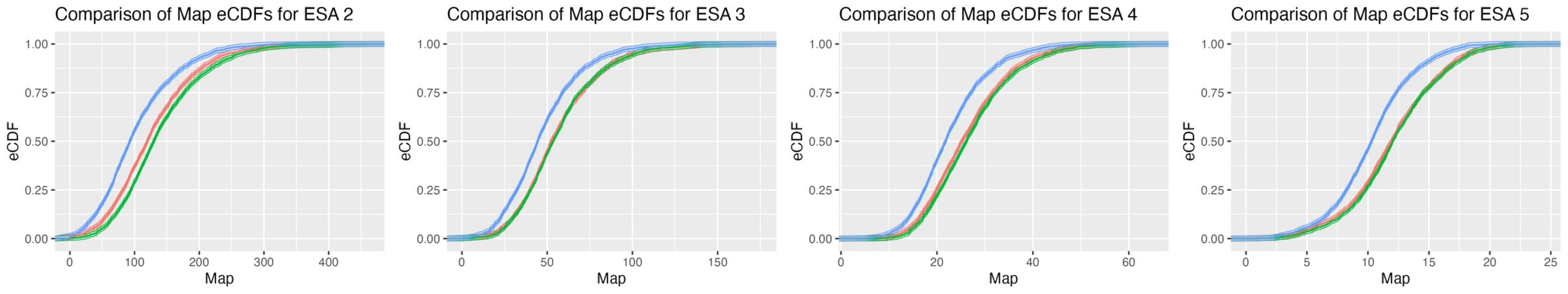}
    \caption{Comparison of eCDF plots with 99\% confidence intervals for 2019B ISOC-like maps resulting from manual culling process, LOTUS Stage 3, and LOTUS Stage 1. Top: eCDFs for ESA step 6. Bottom: eCDFs for ESA steps 2, 3, 4, and 5.}
    \label{Fig:2019B_eCDF}
\end{figure}

We also consider several other measures to determine whether or not LOTUS approximately reproduces the manually culled maps. For example, we can compare the distributions of the fluxes between maps. Row 1 of Figure~\ref{Fig:2019B_eCDF} demonstrates that the distributions of the fluxes for the 2019B ISOC-like map are very similar between the map that results from the manual process and the map that results from LOTUS Stage 3 for ESA 6, and that the 99\% confidence regions even overlap with one another. We see that this is not the case for the LOTUS Stage 1 map. Row 2 displays these empirical cumulative distribution function (eCDF) plots for ESA steps 2, 3, 4, and 5. We see that we again do quite well with ESA steps 3, 4, and 5, but that we tend to struggle with ESA step 2. However, we do note that the eCDF associated with the map generated from the LOTUS Stage 1 labels is much worse for all ESA steps. This is the trend across all maps, ranging from 2009A to 2021B.

We conduct a statistical hypothesis test to evaluate the similarity of these distributions. In particular, we consider a Kolmogorov-Smirnov test \citep{Massey:1951} that considers the following pair of hypotheses
\begin{eqnarray*}
    H_0: F_{SME}(x) = F_{LOTUS}(x) \\
    H_1: F_{SME}(x) \neq F_{LOTUS}(x),
\end{eqnarray*}
where $F(x)$ denotes the eCDF of the fluxes associated with each method (SME and LOTUS). 
We consider this test at the $\alpha=0.01$ significance level, and repeat this test for all ESA steps and all orbit arcs. We find that, over the 26 different maps that consider ESA steps 2 through 6, we fail to reject $H_0$ (indicating that we do not have sufficient evidence to suggest the maps are different from one another) 65\% of the time for LOTUS Stage 3, with a breakdown by ESA step of of 11.54\% for ESA step 2, 76.92\% for ESA step 3, 53.84\% for ESA step 4, 92.31\% for ESA step 5, and 88.46\% for ESA step 6. On the contrary, we fail to reject $H_0$ only 3\% of the time for LOTUS Stage 1, with a breakdown by ESA step of 3.85\% for ESA step 2, 3.85\% for ESA step 3, 7.69\% for ESA step 4, 0.00\% for ESA step 5, and 0.00\% for ESA step 6. 

\begin{figure}[H]
    \centering
    \includegraphics[scale=0.20]{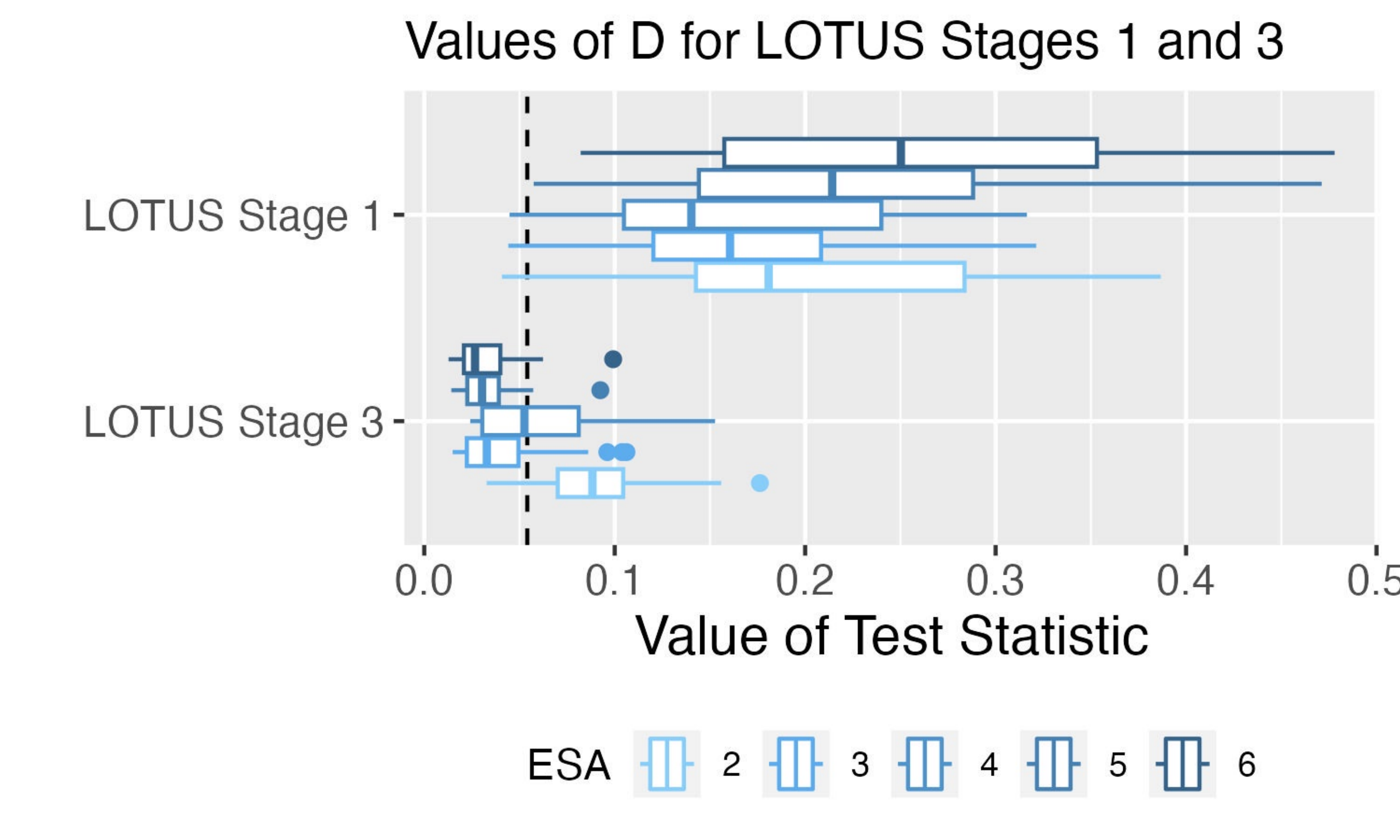}
    \caption{Boxplots of test statistic, $D$, for Kolmogorov-Smirnov test comparing LOTUS eCDFs to SME eCDFs. The dotted vertical line corresponds to the maximum value of the test statistic, $D$, that results in failing to reject $H_0$. Lower values of the test statistic are preferred over larger values, and indicate closer alignment between the two sets of labels.}
    \label{Fig:eCDF_Boxplots_ISOC}
\end{figure}

Figure~\ref{Fig:eCDF_Boxplots_ISOC} portrays the box plots of the test statistic, $D$, broken down by ESA step for pairwise comparisons of the SME eCDFs with LOTUS Stage 1 or 3 eCDFs. We can see from this figure that the LOTUS Stage 3 eCDFs resemble those of the SME much more often than the LOTUS Stage 1 eCDFs.

We can also consider the cross-correlations of the maps within +/- 3 pixels (18 degrees) of a given pixel \citep{Brockwell:1991}. This allows us to determine if the maps are spatially varying consistently with each other. In order to test this, we consider a maximum lag of 3 meaning that pixels in the +/- 3 direction (or +/- 18 degrees) are considered when taking the auto-correlation, which results in a seven-dimensional vector of correlations between the maps. We can then use a t-test to determine whether or not these vectors of cross-correlations are statistically similar \citep{Freund:2010}. For this particular test, the hypotheses are: 
\begin{enumerate}
    \item[$H_0$:] The vector of cross-correlations for the SME generated map with itself is equal to the vector of cross-correlations for the SME generated map with the LOTUS Stage 3 map. 
    \item[$H_1$:] The vector of cross-correlations for the SME generated map with itself is not equal to the vector of cross-correlations for the SME generated map with the LOTUS Stage 3 map. 
\end{enumerate}
When we consider this t-test at the $\alpha=0.01$ significance level, we fail to reject our null hypothesis 100\% of the time for the SME/LOTUS Stage 3 comparisons. This means that for the maps associated with 2019B, we do not have enough evidence to suggest that the cross-correlations between the SME maps and themselves significantly differ from the cross-correlations of the SME maps with the LOTUS Stage 3 maps, at an $\alpha=0.01$ level. This test can also be performed for the LOTUS Stage 1 maps. For these comparisons, we fail to reject the null hypothesis only 20\% of the time. 

\begin{figure}[H]
    \centering
    \includegraphics[scale=0.15]{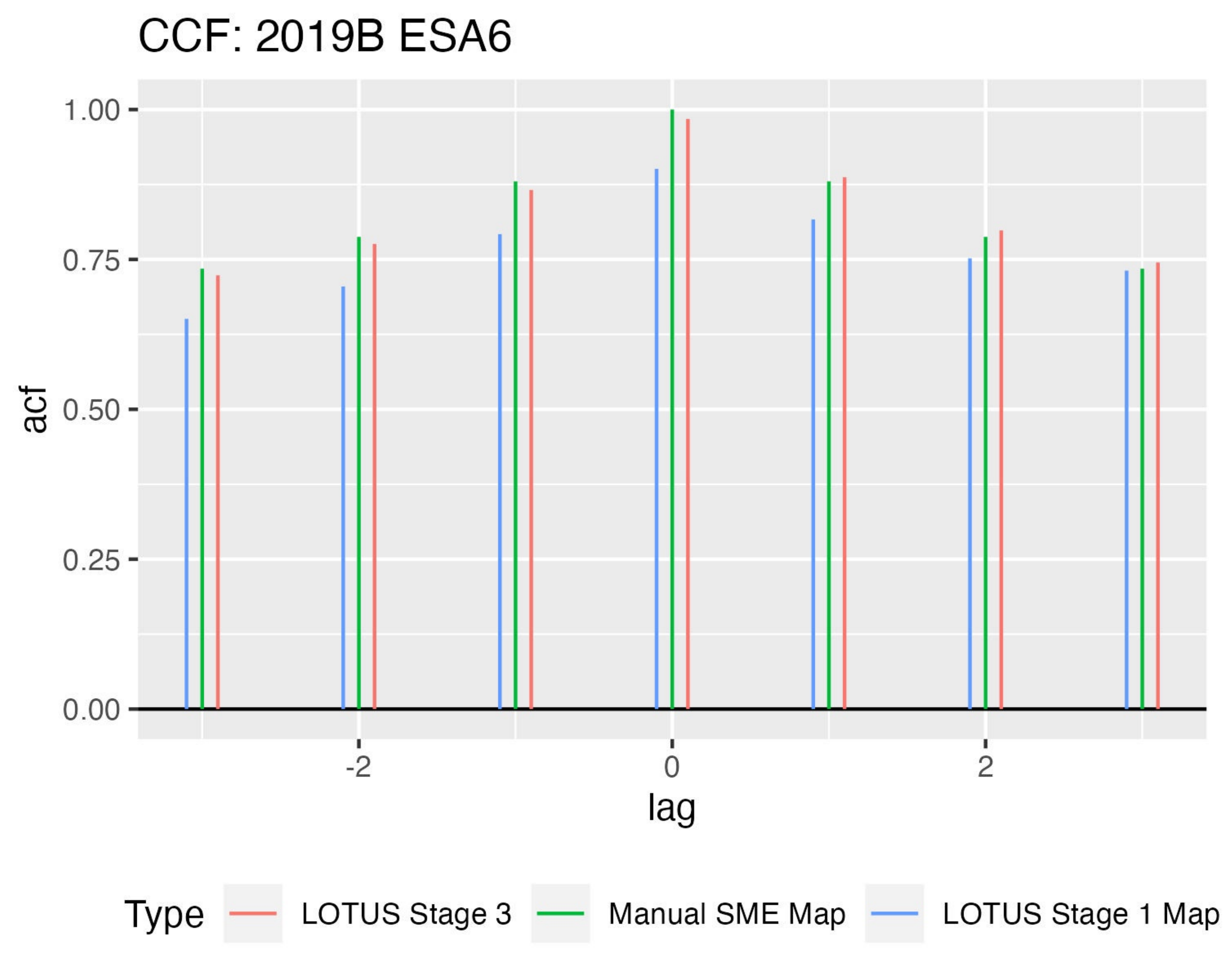} \nonumber \\
    \vspace{3mm}
    \includegraphics[width=\textwidth]{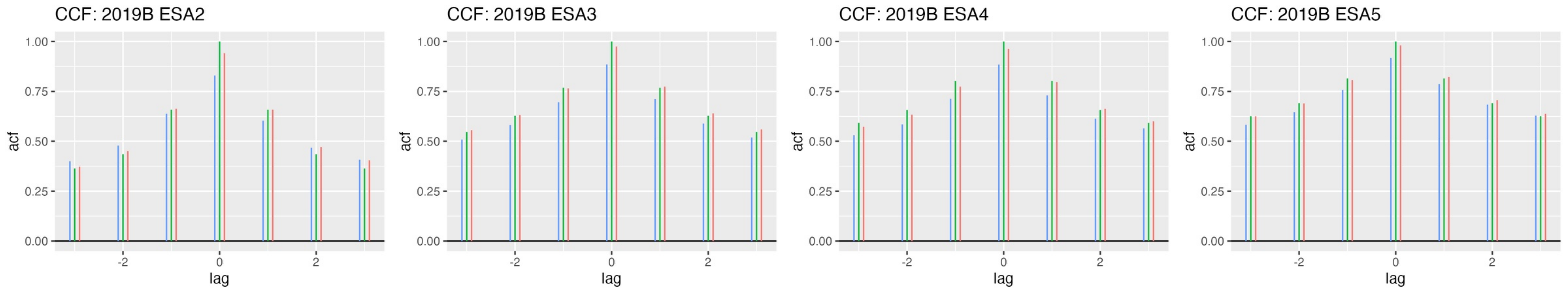}
    \caption{Comparisons of cross-correlations of ISOC-like maps resulting from manual culling process, LOTUS Stage 3, and LOTUS Stage 1 at lags -3, -2, -1, 0, 1, 2, and 3. Top: eCDFs for ESA step 6. Bottom: eCDFs for ESA steps 2, 3, 4, and 5.}
    \label{Fig:2019B_CCF}
\end{figure}

The first row of Figure~\ref{Fig:2019B_CCF} presents the auto-correlation (acf) at each lag (-3, -2, -1, 0, 1, 2, 3) for the 2019B ESA 6 SME map compared against itself (where lag refers to a preceding or following combination of latitude and longitude), the corresponding LOTUS Stage 1 map, and the corresponding LOTUS Stage 3 map. Visually, we can see that the SME generated map is most similar to the LOTUS Stage 3 map. This is the case across all ESAs for map 2019B, which can be seen in row two of Figure~\ref{Fig:2019B_CCF}. 

We can also consider this set of t-tests for all maps ranging from 2009A to 2021B. We fail to reject the null hypothesis for approximately 77.70\% of the LOTUS Stage 3 maps. By ESA step, we fail to reject the null hypothesis for approximately 84.62\% of the maps for ESA step 2, 69.23\% for ESA step 3, 69.23\% for ESA step 4, 80.77\% for ESA step 5, and 84.62\% for ESA step 6. For LOTUS Stage 1, on the other hand, we fail to reject only approximately 45.38\% of the maps, with a per-ESA step break down of 69.23\%, 38.46\%, 42.31\%, 42.31\%, and 34.62\% for ESA steps 2, 3, 4, 5, and 6, respectively.

We can also consider Lin's concordance correlation coefficient \citep{Lin:1989} to determine how well two sets of maps correspond to each other. Like a traditional correlation, Lin's concordance correlation coefficient falls between -1 and 1, and is always less than or equal to the absolute value of Pearson's correlation coefficient. The more ``perfect'' the alignment between two maps, the closer the concordance correlation coefficient will be to 1.00. While Pearson's correlation coefficient considers the correlation that exists between two groups as a whole, Lin's concordance correlation coefficient considers the correlations between individual pairwise observations. Because it considers the correlations at a pairwise level, it is particularly useful for evaluating the agreement between a ``gold'' standard (e.g., maps produced from manually culled SME data) and a proposed replacement (e.g., maps produced from automatically culled LOTUS data).

\begin{figure}[H]
    \centering
    \includegraphics[width=\textwidth]{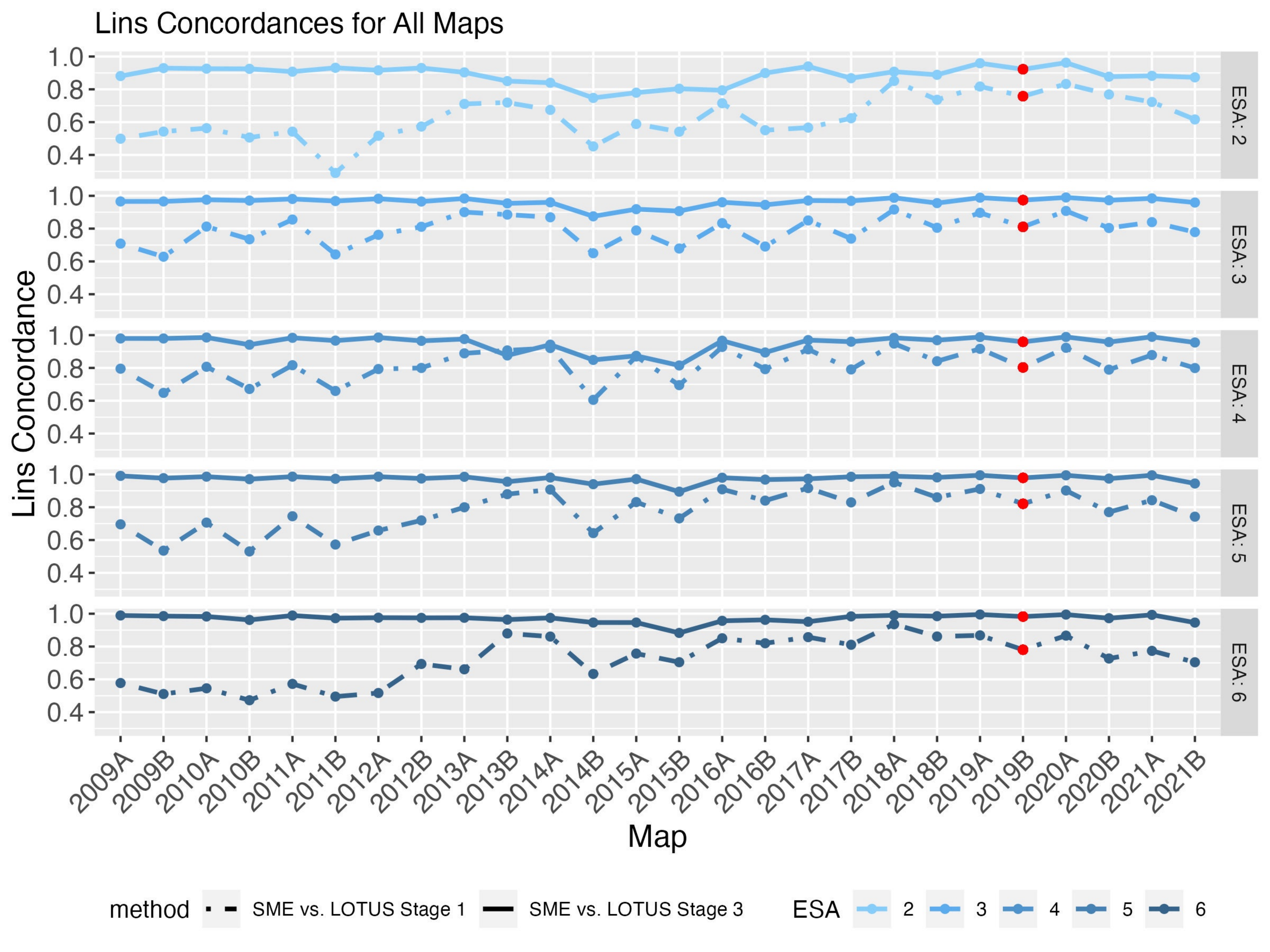}
    \label{SubFig:Lins_1}
    \caption{Lin's Concordance Correlation Coefficients between Manually Labeled SME maps and LOTUS Stage 1 maps (dashed lines), and LOTUS Stage 3 maps (solid lines). Red points correspond to map 2019B.}
    \label{Fig:Lins}
\end{figure}

Figure~\ref{Fig:Lins} portrays the Lin's concordance correlation coefficients between the manual SME maps and LOTUS Stage 1 and LOTUS Stage 3 for Maps 2009A through 2021B for ESAs 2 through 5. From this figure, we can see that the overall output of LOTUS (LOTUS Stage 3) much more closely resembles that of the SME generated maps than does the output of LOTUS Stage 1. The median correlations by ESA step for Stage 1 are 0.60, 0.81, 0.81, 0.81, and 0.74, and for Stage 3 are 0.90, 0.97, 0.97, 0.98, and 0.98. 

We can consider the totality of the results presented in this section to determine whether LOTUS generated maps are a reasonable estimate of SME generated maps. Table~\ref{Tab:MapSuccess} summarizes these results, where green cells correspond to maps whose statistical tests indicate alignment between LOTUS and SME generated maps, while red cells correspond to maps whose statistical tests indicate a lack of alignment between LOTUS and SME generated maps. Depending on how we want to define ``success'' across all tests, we can consider three different options: we can consider that a LOTUS map corresponds to an SME generated map if it passes at least one of the statistical tests (where ``passing'' is indicated by failing to reject $H_0$), if it passes at least two of the statistical tests, or if it passes all three tests (we consider a Bonferonni correction \citep{Rupert:2012} as we consider these three cases). In the first case, we have that 93.08\% of maps pass at least one test; in the second case, we have that 74.61\% of maps pass at least two tests; and in the third case, we have that 52.31\% of maps pass all three tests. We can see that ESA step 2 tends to be the main culprit when it comes to a lack of concordance between SME and LOTUS generated maps. If we exclude ESA step 2 in our calculations, we find that 94.23\% of maps pass at least one test, 88.46\% of maps pass at least two tests, and 64.42\% of maps pass all three tests. We note that our SME judges the reliability of the data used to make these maps. In particular, he considers the group of maps spanning 2015 and 2016 to contain ``concerning'' data, and so it is not necessarily surprising that LOTUS struggles with these maps. We also note that we tend to have particularly good success for ESA steps 5, and 6, and moderate success with ESA steps 3 and 4. Table~\ref{Tab:MapSuccess_byESA} presents the breakdown of results for these three cases across ESA step.

\begin{table}[H]
    \centering
    \begin{tabular}{||m{1.5cm}||m{0.35cm}|m{0.5cm}|m{0.5cm}|m{0.5cm}|m{0.5cm}||m{0.5cm}|m{0.5cm}|m{0.5cm}|m{0.5cm}|m{0.5cm}||m{0.5cm}|m{0.5cm}|m{0.5cm}|m{0.5cm}|m{0.5cm}||}
        \hline
        \rowcolor{myrowcolor!30} Map &  \multicolumn{5}{|c||}{Similar eCDF? (KS Test)}  & \multicolumn{5}{|c||}{Similar CCF? (t Test)} & \multicolumn{5}{|c||}{Lin's Concordance $>0.95$?} \\
        \hline
        \rowcolor{myrowcolor!30} ESA Step & 2 & 3 & 4 & 5 & 6 & 2 & 3 & 4 & 5 & 6 & 2 & 3 & 4 & 5 & 6 \\
        \hline \hline 
        2009A&\cellcolor{red!20} & \cellcolor{red!20} & \cellcolor{green!30} & \cellcolor{green!30} & \cellcolor{green!30} & \cellcolor{green!30} & \cellcolor{red!20} & \cellcolor{red!20} & \cellcolor{red!20} & \cellcolor{red!20} & \cellcolor{red!20} & \cellcolor{green!30} & \cellcolor{green!30} & \cellcolor{green!30} & \cellcolor{green!30} \\
        \hline
        2009B & \cellcolor{green!30} & \cellcolor{green!30} & \cellcolor{green!30} & \cellcolor{green!30} & \cellcolor{green!30} & \cellcolor{green!30} & \cellcolor{red!20} & \cellcolor{red!20} & \cellcolor{green!30} & \cellcolor{green!30} & \cellcolor{red!20} & \cellcolor{green!30} & \cellcolor{green!30} & \cellcolor{green!30} & \cellcolor{green!30} \\
        \hline
        2010A  & \cellcolor{red!20} & \cellcolor{green!30} & \cellcolor{green!30} & \cellcolor{green!30} & \cellcolor{green!30} & \cellcolor{green!30} & \cellcolor{red!20} & \cellcolor{red!20} & \cellcolor{red!20} & \cellcolor{green!30} & \cellcolor{red!20} & \cellcolor{green!30} & \cellcolor{green!30} & \cellcolor{green!30} & \cellcolor{green!30} \\
        \hline
        2010B  & \cellcolor{red!20} & \cellcolor{green!30} & \cellcolor{green!30} & \cellcolor{green!30} & \cellcolor{green!30}  & \cellcolor{green!30} & \cellcolor{red!20} & \cellcolor{red!20} & \cellcolor{red!20} & \cellcolor{red!20} & \cellcolor{red!20} & \cellcolor{green!30} & \cellcolor{red!20} & \cellcolor{green!30} & \cellcolor{green!30} \\
        \hline
        2011A  & \cellcolor{red!20} & \cellcolor{green!30} & \cellcolor{green!30} & \cellcolor{green!30} & \cellcolor{green!30} & \cellcolor{green!30} & \cellcolor{red!20}  & \cellcolor{red!20} & \cellcolor{red!20} & \cellcolor{green!30} & \cellcolor{red!20} & \cellcolor{green!30} & \cellcolor{green!30} & \cellcolor{green!30} & \cellcolor{green!30} \\
        \hline
        2011B & \cellcolor{green!30} & \cellcolor{green!30} & \cellcolor{green!30} & \cellcolor{green!30} & \cellcolor{green!30} & \cellcolor{green!30} & \cellcolor{green!30} & \cellcolor{green!30} & \cellcolor{green!30} & \cellcolor{green!30} & \cellcolor{red!20} & \cellcolor{green!30} & \cellcolor{green!30} & \cellcolor{green!30} & \cellcolor{green!30} \\
        \hline
        2012A  & \cellcolor{red!20} & \cellcolor{green!30} & \cellcolor{green!30} & \cellcolor{green!30} & \cellcolor{green!30} & \cellcolor{green!30} & \cellcolor{green!30} & \cellcolor{green!30} & \cellcolor{green!30} & \cellcolor{green!30} & \cellcolor{red!20} & \cellcolor{green!30} & \cellcolor{green!30} & \cellcolor{green!30} & \cellcolor{green!30} \\
        \hline
        2012B  & \cellcolor{red!20} & \cellcolor{green!30} & \cellcolor{green!30} & \cellcolor{green!30} & \cellcolor{green!30} & \cellcolor{green!30} & \cellcolor{green!30} & \cellcolor{green!30} & \cellcolor{green!30} & \cellcolor{green!30} & \cellcolor{red!20} & \cellcolor{green!30} & \cellcolor{green!30} & \cellcolor{green!30} & \cellcolor{green!30} \\
        \hline
        2013A  & \cellcolor{red!20} & \cellcolor{green!30} & \cellcolor{green!30} & \cellcolor{green!30} & \cellcolor{green!30} & \cellcolor{green!30} & \cellcolor{green!30} & \cellcolor{green!30} & \cellcolor{green!30} & \cellcolor{green!30} & \cellcolor{red!20} & \cellcolor{green!30} & \cellcolor{green!30} & \cellcolor{green!30} & \cellcolor{green!30} \\
        \hline
        2013B  & \cellcolor{red!20} & \cellcolor{red!20} & \cellcolor{red!20} & \cellcolor{green!30} & \cellcolor{green!30} & \cellcolor{green!30} & \cellcolor{green!30} & \cellcolor{green!30} & \cellcolor{green!30} & \cellcolor{green!30} & \cellcolor{red!20} & \cellcolor{green!30} & \cellcolor{red!20} & \cellcolor{green!30} & \cellcolor{green!30} \\
        \hline
        2014A  & \cellcolor{red!20}  & \cellcolor{red!20} & \cellcolor{green!30} & \cellcolor{green!30} & \cellcolor{green!30} & \cellcolor{green!30} & \cellcolor{green!30} & \cellcolor{green!30} & \cellcolor{green!30} & \cellcolor{green!30} & \cellcolor{red!20} & \cellcolor{green!30} & \cellcolor{red!20} & \cellcolor{green!30} & \cellcolor{green!30} \\
        \hline
        2014B  & \cellcolor{red!20} & \cellcolor{red!20} & \cellcolor{red!20} & \cellcolor{green!30} & \cellcolor{green!30} & \cellcolor{green!30} & \cellcolor{green!30} & \cellcolor{green!30} & \cellcolor{green!30} & \cellcolor{green!30} & \cellcolor{red!20} & \cellcolor{red!20} & \cellcolor{red!20} & \cellcolor{green!30} & \cellcolor{green!30} \\
        \hline
        2015A  & \cellcolor{red!20} & \cellcolor{red!20} & \cellcolor{red!20} & \cellcolor{green!30} & \cellcolor{green!30}  & \cellcolor{red!20} & \cellcolor{red!20} & \cellcolor{red!20} & \cellcolor{green!30} & \cellcolor{green!30} & \cellcolor{red!20} & \cellcolor{red!20} & \cellcolor{red!20} & \cellcolor{red!20} & \cellcolor{green!30} \\
        \hline
        2015B  & \cellcolor{red!20} & \cellcolor{red!20} & \cellcolor{red!20}  & \cellcolor{red!20} & \cellcolor{red!20} & \cellcolor{green!30} & \cellcolor{red!20} & \cellcolor{red!20} & \cellcolor{red!20} & \cellcolor{red!20} & \cellcolor{red!20}  & \cellcolor{red!20} & \cellcolor{red!20} & \cellcolor{red!20} & \cellcolor{red!20} \\
        \hline
        2016A  & \cellcolor{red!20} & \cellcolor{green!30} & \cellcolor{red!20} & \cellcolor{green!30} & \cellcolor{red!20} & \cellcolor{red!20} & \cellcolor{red!20} & \cellcolor{green!30} & \cellcolor{green!30} & \cellcolor{green!30} & \cellcolor{red!20} & \cellcolor{green!30} & \cellcolor{green!30} & \cellcolor{green!30} & \cellcolor{green!30} \\
        \hline
        2016B  & \cellcolor{red!20} & \cellcolor{green!30} & \cellcolor{red!20} & \cellcolor{green!30} & \cellcolor{green!30} & \cellcolor{green!30} & \cellcolor{green!30} & \cellcolor{green!30} & \cellcolor{green!30} & \cellcolor{green!30} & \cellcolor{red!20} & \cellcolor{green!30} & \cellcolor{red!20} & \cellcolor{green!30} & \cellcolor{green!30} \\
        \hline
        2017A  & \cellcolor{green!30} & \cellcolor{green!30} & \cellcolor{red!20} & \cellcolor{green!30} & \cellcolor{green!30} & \cellcolor{green!30} & \cellcolor{green!30} & \cellcolor{green!30} & \cellcolor{green!30} & \cellcolor{green!30} & \cellcolor{red!20}  & \cellcolor{green!30} & \cellcolor{green!30}  & \cellcolor{green!30} & \cellcolor{green!30} \\
        \hline
        2017B  & \cellcolor{red!20} & \cellcolor{green!30} & \cellcolor{red!20} & \cellcolor{green!30} & \cellcolor{green!30} & \cellcolor{green!30} & \cellcolor{green!30} & \cellcolor{green!30} & \cellcolor{green!30} & \cellcolor{green!30} & \cellcolor{red!20} & \cellcolor{green!30} & \cellcolor{green!30} & \cellcolor{green!30} & \cellcolor{green!30} \\
        \hline
        2018A  & \cellcolor{red!20} & \cellcolor{green!30} & \cellcolor{red!20} & \cellcolor{green!30} & \cellcolor{green!30} & \cellcolor{green!30} & \cellcolor{green!30} & \cellcolor{green!30} & \cellcolor{green!30} & \cellcolor{green!30} & \cellcolor{red!20} & \cellcolor{green!30} & \cellcolor{green!30} & \cellcolor{green!30} & \cellcolor{green!30} \\
        \hline
        2018B  & \cellcolor{red!20} & \cellcolor{green!30} & \cellcolor{green!30} & \cellcolor{green!30} & \cellcolor{green!30} & \cellcolor{green!30} & \cellcolor{green!30} & \cellcolor{green!30} & \cellcolor{green!30} & \cellcolor{green!30} & \cellcolor{red!20} & \cellcolor{green!30} & \cellcolor{green!30} & \cellcolor{green!30} & \cellcolor{green!30} \\
        \hline
        2019A  & \cellcolor{red!20} & \cellcolor{green!30} & \cellcolor{green!30} & \cellcolor{green!30} & \cellcolor{green!30} & \cellcolor{green!30} & \cellcolor{green!30} & \cellcolor{green!30} & \cellcolor{green!30} & \cellcolor{green!30} & \cellcolor{green!30} & \cellcolor{green!30} & \cellcolor{green!30} & \cellcolor{green!30} & \cellcolor{green!30} \\
        \hline
        2019B  & \cellcolor{red!20} & \cellcolor{green!30} & \cellcolor{green!30} & \cellcolor{green!30} & \cellcolor{green!30} & \cellcolor{green!30} & \cellcolor{green!30} & \cellcolor{green!30} & \cellcolor{green!30} & \cellcolor{green!30} & \cellcolor{red!20} & \cellcolor{green!30} & \cellcolor{green!30} & \cellcolor{green!30} & \cellcolor{green!30} \\
        \hline
        2020A & \cellcolor{green!30} & \cellcolor{green!30} & \cellcolor{green!30} & \cellcolor{green!30} & \cellcolor{green!30} & \cellcolor{green!30} & \cellcolor{green!30} & \cellcolor{green!30} & \cellcolor{green!30} & \cellcolor{green!30} & \cellcolor{green!30} & \cellcolor{green!30} & \cellcolor{green!30} & \cellcolor{green!30} & \cellcolor{green!30} \\
        \hline
        2020B & \cellcolor{red!20} & \cellcolor{green!30} & \cellcolor{green!30} & \cellcolor{green!30} & \cellcolor{green!30} & \cellcolor{green!30} & \cellcolor{green!30} & \cellcolor{green!30} & \cellcolor{green!30} & \cellcolor{green!30} & \cellcolor{red!20} & \cellcolor{green!30} & \cellcolor{green!30} & \cellcolor{green!30} & \cellcolor{green!30} \\
        \hline
        2021A  & \cellcolor{red!20} & \cellcolor{green!30} & \cellcolor{green!30} & \cellcolor{green!30} & \cellcolor{green!30} & \cellcolor{green!30} & \cellcolor{green!30} & \cellcolor{green!30} & \cellcolor{green!30} & \cellcolor{green!30} & \cellcolor{red!20} & \cellcolor{green!30} & \cellcolor{green!30} & \cellcolor{green!30} & \cellcolor{green!30} \\
        \hline
        2021B  & \cellcolor{red!20} & \cellcolor{green!30}  & \cellcolor{red!20} & \cellcolor{green!30} & \cellcolor{green!30} & \cellcolor{green!30} & \cellcolor{green!30} & \cellcolor{green!30} & \cellcolor{green!30} & \cellcolor{green!30} & \cellcolor{red!20} & \cellcolor{green!30} & \cellcolor{green!30} & \cellcolor{green!30} & \cellcolor{green!30} \\
        \hline
    \end{tabular}
    \caption{Breakdown of Test Success by Map and ESA step. Green cells correspond to maps whose tests indicate concordance between LOTUS and SME generated maps. Red cells correspond to maps whose tests indicate lack of concordance between LOTUS and SME generated maps.}
    \label{Tab:MapSuccess}
\end{table}

\begin{table}[H]
    \centering
    \begin{tabular}{|c||c|c|c|c|c|}
        \hline
         \rowcolor{myrowcolor!30} ESA Step & 2 & 3 & 4 & 5 & 6 \\
         \hline \hline
         Case 1 & 88.5\% & 92.3\% & 92.3\% & 96.2\% & 96.2\% \\
         \hline
         Case 2 & 19.2\% & 84.6\% & 76.9\% & 96.2\% & 96.2\% \\
         \hline
         Case 3 &  3.9\% & 57.7\% & 38.5\% & 76.9\% & 84.6\% \\
         \hline 
    \end{tabular}
    \caption{Summary of Table~\ref{Tab:MapSuccess}. Percentage of maps that pass at least one (Case 1), two (Case 2), and three (Case 3) of the three considered statistical tests, by ESA step.}
    \label{Tab:MapSuccess_byESA}
\end{table}

\section{Conclusion}                                                     %
\label{Sec:Conclusion}                                                   %
In this paper, we discuss LOTUS, an automated process for distinguishing between energetic neutral atoms and incidental background particles for data generated by NASA's Interstellar Boundary Explorer Mission. LOTUS offers a more objective, rapid alternative compared to the current manual process. This paper demonstrates that LOTUS is capable of reproducing ENA rates and maps that are relatively comparable to those that result from the manual culling process. We purport that LOTUS is useful for reproducing maps for ESA steps 5 and 6, and that LOTUS provides a useful and time-effective starting point for ESA steps 3 and 4. In either case, LOTUS stands as a convenient tool that will not only save an SME time in distinguishing between ``good times'' and ``bad times'', and also offers a standardized process for obtaining labels.

Automated methods such as LOTUS can also provide faster data processing and map-making for future missions, such as the Interstellar Mapping and Acceleration Probe (IMAP)~\citep{McComas:2018}, once training data becomes available for this mission. As IMAP makes new observations of the ribbon, its data quality can be characterized more rapidly, compared to IBEX, and improvements over past observations demonstrated more promptly than manual methods, of time-sensitive value as the heliospheric community contemplates missions such as NASA's Interstellar Probe~\citep{McNutt:2022}. The IBEX and IMAP data selection process can also be cross-validated, to ensure similar measures are used across both missions, simplifying efforts to fuse the data sets. Finally, the automated process offers the potential to tune the selection parameters, if justified. This potential may enable more complete utilization of ENA data and more accurate studies of the outer heliosphere.

LOTUS provides a noteworthy stepping stone towards automating the time-intensive manual culling process. While it may take upwards of several weeks to manually label a new orbit of data, LOTUS is able to generate labels in a matter of minutes, at minimum providing SME with a valuable standardized resource that can facilitate the manual culling process.
\newpage

\bibliographystyle{ieeetr}
\bibliography{0_LOTUS.bbl}

\begin{thebibliography}{10}

\bibitem{Funsten:2009a}
H.~Funsten, F.~Allegrini, P.~Bochsler, G.~Dunn, S.~Ellis, D.~Everett, M.~Fagan,
  S.~Fuselier, M.~Granoff, M.~Gruntman, A.~Guthrie, J.~Hanley, R.~Harper,
  D.~Heirtzler, P.~Janzen, K.~Kihara, B.~King, H.~Kucharek, M.~Manzo, M.~Maple,
  K.~Mashburn, D.~McComas, E.~Moebius, J.~Nolin, D.~Piazza, S.~Pope,
  D.~Reisenfeld, B.~Rodriguez, E.~Roelof, L.~Saul, S.~Turco, P.~Valek,
  S.~Weidner, P.~Wurz, and S.~Zaffke, ``The interstellar boundary explorer high
  energy (ibex-hi) neutral atom imager,'' {\em Space Science Reviews},
  vol.~146, no.~1, pp.~75--103, 2009.

\bibitem{McComas:2009}
D.~McComas, F.~Allegrini, P.~Bochsler, M.~Bzowski, E.~Christian, G.~Crew,
  R.~DeMajistre, H.~Fahr, H.~Fichtner, P.~Frisch, H.~Funsten, S.~Fuselier,
  G.~Gloeckler, M.~Gruntman, J.~Heerikhuisen, V.~Izmodenov, P.~Janzen,
  P.~Knappenberger, S.~Krimigis, H.~Kucharek, M.~Lee, G.~Livadiotis, S.~Livi,
  R.~Macdowall, D.~Mitchell, E.~M{\"o}bius, T.~Moore, N.~Pogorelov,
  D.~Reisenfeld, E.~Roelof, L.~Saul, N.~Schwadron, P.~Valek, R.~Vanderspek,
  P.~Wurz, and G.~Zank, ``Global observations of the interstellar interaction
  from the interstellar boundary explorer (ibex),'' {\em Science}, vol.~326,
  no.~5955, pp.~959--962, 2009.

\bibitem{Reisenfeld:2021}
D.~B. Reisenfeld, M.~Bzowski, H.~O. Funsten, J.~Heerikhuisen, P.~H. Janzen,
  M.~A. Kubiak, D.~J. McComas, N.~A. Schwadron, J.~M. Sokół, A.~Zimorino, and
  E.~J. Zirnstein, ``A three-dimensional map of the heliosphere from ibex,''
  {\em The Astrophysical Journal Supplement Series}, vol.~254, p.~40, jun 2021.

\bibitem{Osthus:2023}
D.~Osthus, L.~J.~B. Brian P.~Weaver, K.~R. Moran, M.~A. Stricklin, E.~J.
  Zirnstein, P.~H. Janzen, and D.~B. Reisenfeld, ``Towards improved
  heliospheric sky map estimation with theseus,'' {\em Technometrics}, vol.~0,
  no.~ja, pp.~1--27, 2023.

\bibitem{Funsten:2009b}
H.~Funsten, F.~Allegrini, G.~Crew, R.~DeMajistre, P.~Frisch, S.~Fuselier,
  M.~Gruntman, P.~Janzen, D.~McComas, E.~M{\"o}bius, B.~Randol, D.~Reisenfeld,
  E.~Roelof, and N.~Schwadron, ``Structures and spectral variations of the
  outer heliosphere in ibex energetic neutral atom maps,'' {\em Space Sci.
  Rev}, vol.~326, p.~2021, 2009.

\bibitem{McComas:2018}
D.~J. McComas, E.~R. Christian, N.~A. Schwadron, N.~Fox, J.~Westlake,
  F.~Allegrini, D.~N. Baker, D.~Biesecker, M.~Bzowski, G.~Clark, C.~M.~S.
  Cohen, I.~Cohen, M.~A. Dayeh, R.~Decker, G.~A. de~Nolfo, D.~M. I., R.~W.
  Ebert, H.~A. Elliott, H.~Fahr, P.~C. Frisch, H.~O. Funsten, S.~A. Fuselier,
  A.~Galli, A.~B. Galvin, J.~Giacalone, M.~Gkioulidou, F.~Guo, M.~Horanyi,
  P.~Isenberg, P.~Janzen, L.~M. Kistler, K.~Korreck, M.~A. Kubiak, H.~Kucharek,
  B.~A. Larsen, R.~A. Leske, N.~Lugaz, J.~Luhmann, W.~Matthaeus, D.~Mitchell,
  E.~Moebius, K.~Ogasawara, D.~B. Reisenfeld, J.~D. Richardson, C.~T. Russell,
  J.~M. Sok{\'o'}l, H.~E. Spence, R.~Skoug, Z.~Sternovsky, P.~Swaczyna, M.~E.
  Wiedenbeck, P.~Wurz, G.~P. Zank, and E.~J. Zirnstein, ``Interstellar mapping
  and acceleration probe (imap): A new nasa mission,'' {\em Space Science
  Reviews}, vol.~214, p.~116, October 2018.

\bibitem{Allegrini:2009}
F.~Allegrini, G.~Crew, D.~Demkee, H.~Funsten, D.~McComas, B.~Randol,
  B.~Rodriguez, N.~Schwadron, P.~Valek, and S.~Weidner, ``The ibex background
  monitor,'' {\em Space Science Reviews}, vol.~146, pp.~105--115, 2008.

\bibitem{Acton:1996}
C.~H. Acton, ``Ancillary data services of nasa's navigation and ancillary
  information facility,'' {\em Planetary and Space Science}, vol.~44, no.~1,
  pp.~65--70, 1996.

\bibitem{Acton:2018}
C.~Acton, N.~Bachman, B.~Semenov, and E.~Wright, ``A look towards the future in
  the handling of space science mission geometry,'' {\em Planetary and Space
  Science}, vol.~150, pp.~9--12, 2018.

\bibitem{Hastie:2009}
T.~Hastie, R.~Tibshirani, J.~H. Friedman, and J.~H. Friedman, {\em The elements
  of statistical learning: data mining, inference, and prediction}, vol.~2.
\newblock Springer, 2009.

\bibitem{Bishop:2006}
C.~M. Bishop, {\em Pattern recognition and machine learning}.
\newblock Springer, 1~ed., 2006.

\bibitem{R:2021}
{R Core Team}, {\em R: A Language and Environment for Statistical Computing}.
\newblock R Foundation for Statistical Computing, Vienna, Austria, 2021.

\bibitem{ranger:2017}
M.~N. Wright and A.~Ziegler, ``{ranger}: A fast implementation of random
  forests for high dimensional data in {C++} and {R},'' {\em Journal of
  Statistical Software}, vol.~77, no.~1, pp.~1--17, 2017.

\bibitem{Venabls:2013}
W.~N. Venables and B.~D. Ripley, {\em Modern applied statistics with S-PLUS}.
\newblock Springer Science \& Business Media, 2013.

\bibitem{Massey:1951}
F.~J. Massey~Jr, ``The kolmogorov-smirnov test for goodness of fit,'' {\em
  Journal of the American statistical Association}, vol.~46, no.~253,
  pp.~68--78, 1951.

\bibitem{Brockwell:1991}
P.~J. Brockwell and R.~A. Davis, {\em Time series: theory and methods}.
\newblock Springer science \& business media, 1991.

\bibitem{Freund:2010}
R.~J. Freund, W.~J. Wilson, and D.~L. Mohr, {\em Statistical methods, students
  solutions manual (e-only)}.
\newblock Academic Press, 2010.

\bibitem{Lin:1989}
L.~I.-K. Lin, ``A concordance correlation coefficient to evaluate
  reproducibility,'' {\em Biometrics}, no.~1, pp.~255--268, 1989.

\bibitem{Rupert:2012}
G.~Rupert~Jr, ``Simultaneous statistical inference,'' 2012.

\bibitem{McNutt:2022}
R.~L. McNutt~Jr, R.~F. Wimmer-Schweingruber, M.~Gruntman, S.~M. Krimigis, E.~C.
  Roelof, P.~C. Brandt, S.~R. Vernon, M.~V. Paul, R.~W. Stough, and J.~D.
  Kinnison, ``Interstellar probe--destination: universe!,'' {\em Acta
  Astronautica}, vol.~196, pp.~13--28, 2022.

\bibitem{Annex2020}
A.~M. Annex, B.~Pearson, B.~Seignovert, B.~T. Carcich, H.~Eichhorn, J.~A.
  Mapel, J.~L.~F. von Forstner, J.~McAuliffe, J.~D. del Rio, K.~L. Berry, K.-M.
  Aye, M.~Stefko, M.~de~Val-Borro, S.~Kulumani, and S.~ya~Murakami, ``Spiceypy:
  a pythonic wrapper for the spice toolkit,'' {\em Journal of Open Source
  Software}, vol.~5, no.~46, p.~2050, 2020.

\bibitem{astropy:2013}
{Astropy Collaboration}, T.~P. {Robitaille}, E.~J. {Tollerud}, P.~{Greenfield},
  M.~{Droettboom}, E.~{Bray}, T.~{Aldcroft}, M.~{Davis}, A.~{Ginsburg}, A.~M.
  {Price-Whelan}, W.~E. {Kerzendorf}, A.~{Conley}, N.~{Crighton}, K.~{Barbary},
  D.~{Muna}, H.~{Ferguson}, F.~{Grollier}, M.~M. {Parikh}, P.~H. {Nair}, H.~M.
  {Unther}, C.~{Deil}, J.~{Woillez}, S.~{Conseil}, R.~{Kramer}, J.~E.~H.
  {Turner}, L.~{Singer}, R.~{Fox}, B.~A. {Weaver}, V.~{Zabalza}, Z.~I.
  {Edwards}, K.~{Azalee Bostroem}, D.~J. {Burke}, A.~R. {Casey}, S.~M.
  {Crawford}, N.~{Dencheva}, J.~{Ely}, T.~{Jenness}, K.~{Labrie}, P.~L. {Lim},
  F.~{Pierfederici}, A.~{Pontzen}, A.~{Ptak}, B.~{Refsdal}, M.~{Servillat}, and
  O.~{Streicher}, ``Astropy: A community python package for astronomy,'' {\em
  Astronomy and Astrophysics}, vol.~558, p.~A33, 2013.

\bibitem{astropy:2018}
{Astropy Collaboration}, A.~M. {Price-Whelan}, B.~M. {Sip{\H{o}}cz}, H.~M.
  {G{\"u}nther}, P.~L. {Lim}, S.~M. {Crawford}, S.~{Conseil}, D.~L. {Shupe},
  M.~W. {Craig}, N.~{Dencheva}, A.~{Ginsburg}, J.~T. {VanderPlas}, L.~D.
  {Bradley}, D.~{P{\'e}rez-Su{\'a}rez}, M.~{de Val-Borro}, T.~L. {Aldcroft},
  K.~L. {Cruz}, T.~P. {Robitaille}, E.~J. {Tollerud}, C.~{Ardelean},
  T.~{Babej}, Y.~P. {Bach}, M.~{Bachetti}, A.~V. {Bakanov}, S.~P. {Bamford},
  G.~{Barentsen}, P.~{Barmby}, A.~{Baumbach}, K.~L. {Berry}, F.~{Biscani},
  M.~{Boquien}, K.~A. {Bostroem}, L.~G. {Bouma}, G.~B. {Brammer}, E.~M. {Bray},
  H.~{Breytenbach}, H.~{Buddelmeijer}, D.~J. {Burke}, G.~{Calderone}, J.~L.
  {Cano Rodr{\'\i}guez}, M.~{Cara}, J.~V.~M. {Cardoso}, S.~{Cheedella},
  Y.~{Copin}, L.~{Corrales}, D.~{Crichton}, D.~{D'Avella}, C.~{Deil},
  {\'E}.~{Depagne}, J.~P. {Dietrich}, A.~{Donath}, M.~{Droettboom}, N.~{Earl},
  T.~{Erben}, S.~{Fabbro}, L.~A. {Ferreira}, T.~{Finethy}, R.~T. {Fox}, L.~H.
  {Garrison}, S.~L.~J. {Gibbons}, D.~A. {Goldstein}, R.~{Gommers}, J.~P.
  {Greco}, P.~{Greenfield}, A.~M. {Groener}, F.~{Grollier}, A.~{Hagen},
  P.~{Hirst}, D.~{Homeier}, A.~J. {Horton}, G.~{Hosseinzadeh}, L.~{Hu}, J.~S.
  {Hunkeler}, {\v{Z}}.~{Ivezi{\'c}}, A.~{Jain}, T.~{Jenness}, G.~{Kanarek},
  S.~{Kendrew}, N.~S. {Kern}, W.~E. {Kerzendorf}, A.~{Khvalko}, J.~{King},
  D.~{Kirkby}, A.~M. {Kulkarni}, A.~{Kumar}, A.~{Lee}, D.~{Lenz}, S.~P.
  {Littlefair}, Z.~{Ma}, D.~M. {Macleod}, M.~{Mastropietro}, C.~{McCully},
  S.~{Montagnac}, B.~M. {Morris}, M.~{Mueller}, S.~J. {Mumford}, D.~{Muna},
  N.~A. {Murphy}, S.~{Nelson}, G.~H. {Nguyen}, J.~P. {Ninan}, M.~{N{\"o}the},
  S.~{Ogaz}, S.~{Oh}, J.~K. {Parejko}, N.~{Parley}, S.~{Pascual}, R.~{Patil},
  A.~A. {Patil}, A.~L. {Plunkett}, J.~X. {Prochaska}, T.~{Rastogi}, V.~{Reddy
  Janga}, J.~{Sabater}, P.~{Sakurikar}, M.~{Seifert}, L.~E. {Sherbert},
  H.~{Sherwood-Taylor}, A.~Y. {Shih}, J.~{Sick}, M.~T. {Silbiger},
  S.~{Singanamalla}, L.~P. {Singer}, P.~H. {Sladen}, K.~A. {Sooley},
  S.~{Sornarajah}, O.~{Streicher}, P.~{Teuben}, S.~W. {Thomas}, G.~R.
  {Tremblay}, J.~E.~H. {Turner}, V.~{Terr{\'o}n}, M.~H. {van Kerkwijk}, A.~{de
  la Vega}, L.~L. {Watkins}, B.~A. {Weaver}, J.~B. {Whitmore}, J.~{Woillez},
  V.~{Zabalza}, and {Astropy Contributors}, ``{The Astropy Project: Building an
  Open-science Project and Status of the v2.0 Core Package},'' {\em The
  Astrophysical Journal}, vol.~156, no.~3, p.~123, 2018.

\bibitem{astropy:2022}
{Astropy Collaboration}, A.~M. {Price-Whelan}, P.~L. {Lim}, N.~{Earl},
  N.~{Starkman}, L.~{Bradley}, D.~L. {Shupe}, A.~A. {Patil}, L.~{Corrales},
  C.~E. {Brasseur}, M.~{N{"o}the}, A.~{Donath}, E.~{Tollerud}, B.~M. {Morris},
  A.~{Ginsburg}, E.~{Vaher}, B.~A. {Weaver}, J.~{Tocknell}, W.~{Jamieson},
  M.~H. {van Kerkwijk}, T.~P. {Robitaille}, B.~{Merry}, M.~{Bachetti}, H.~M.
  {G{"u}nther}, T.~L. {Aldcroft}, J.~A. {Alvarado-Montes}, A.~M. {Archibald},
  A.~{B{'o}di}, S.~{Bapat}, G.~{Barentsen}, J.~{Baz{'a}n}, M.~{Biswas},
  M.~{Boquien}, D.~J. {Burke}, D.~{Cara}, M.~{Cara}, K.~E. {Conroy},
  S.~{Conseil}, M.~W. {Craig}, R.~M. {Cross}, K.~L. {Cruz}, F.~{D'Eugenio},
  N.~{Dencheva}, H.~A.~R. {Devillepoix}, J.~P. {Dietrich}, A.~D. {Eigenbrot},
  T.~{Erben}, L.~{Ferreira}, D.~{Foreman-Mackey}, R.~{Fox}, N.~{Freij},
  S.~{Garg}, R.~{Geda}, L.~{Glattly}, Y.~{Gondhalekar}, K.~D. {Gordon},
  D.~{Grant}, P.~{Greenfield}, A.~M. {Groener}, S.~{Guest}, S.~{Gurovich},
  R.~{Handberg}, A.~{Hart}, Z.~{Hatfield-Dodds}, D.~{Homeier},
  G.~{Hosseinzadeh}, T.~{Jenness}, C.~K. {Jones}, P.~{Joseph}, J.~B.
  {Kalmbach}, E.~{Karamehmetoglu}, M.~{Ka{l}uszy{'n}ski}, M.~S.~P. {Kelley},
  N.~{Kern}, W.~E. {Kerzendorf}, E.~W. {Koch}, S.~{Kulumani}, A.~{Lee},
  C.~{Ly}, Z.~{Ma}, C.~{MacBride}, J.~M. {Maljaars}, D.~{Muna}, N.~A. {Murphy},
  H.~{Norman}, R.~{O'Steen}, K.~A. {Oman}, C.~{Pacifici}, S.~{Pascual},
  J.~{Pascual-Granado}, R.~R. {Patil}, G.~I. {Perren}, T.~E. {Pickering},
  T.~{Rastogi}, B.~R. {Roulston}, D.~F. {Ryan}, E.~S. {Rykoff}, J.~{Sabater},
  P.~{Sakurikar}, J.~{Salgado}, A.~{Sanghi}, N.~{Saunders}, V.~{Savchenko},
  L.~{Schwardt}, M.~{Seifert-Eckert}, A.~Y. {Shih}, A.~S. {Jain}, G.~{Shukla},
  J.~{Sick}, C.~{Simpson}, S.~{Singanamalla}, L.~P. {Singer}, J.~{Singhal},
  M.~{Sinha}, B.~M. {Sip{H{o}}cz}, L.~R. {Spitler}, D.~{Stansby},
  O.~{Streicher}, J.~{{S}umak}, J.~D. {Swinbank}, D.~S. {Taranu}, N.~{Tewary},
  G.~R. {Tremblay}, M.~d. {Val-Borro}, S.~J. {Van Kooten}, Z.~{Vasovi{'c}},
  S.~{Verma}, J.~V. {de Miranda Cardoso}, P.~K.~G. {Williams}, T.~J. {Wilson},
  B.~{Winkel}, W.~M. {Wood-Vasey}, R.~{Xue}, P.~{Yoachim}, C.~{Zhang},
  A.~{Zonca}, and {Astropy Project Contributors}, ``{The Astropy Project:
  Sustaining and Growing a Community-oriented Open-source Project and the
  Latest Major Release (v5.0) of the Core Package},'' {\em The Astrophysical
  Journal}, vol.~935, no.~2, p.~167, 2022.

\end{thebibliography}


\clearpage

\appendix
\section*{Supplemental Material}

\section{SPICE Overview}                                                 %
\label{App:SpiceSpiceBaby}                                               %
Data selection for the Interstellar Boundary Explorer mission has, since 2008, relied on a human curator to assess spacecraft orientation. When oriented toward nearby objects, such as the Sun, Moon, or Earth, the IBEX energetic neutral atom detector can reach much higher noise levels than typical for heliospheric observations. Although maneuvers are designed to minimize exposure to the Sun, there are some periods when IBEX is oriented toward the Moon or the Earth's magnetosphere. Automated data selection benefits from knowing these periods. The spacecraft orientation can be deterministically calculated for past orbits, although on-demand orbital maneuvers mean that it is not generally predictable into the future. Because the onset and cessation of high-noise periods due to IBEX orientation typically manifest as abrupt changes to a different ENA background level, which can be difficult for many machine-learning methods to replicate, the authors anticipated that incorporating orientation information into the learning process would increase training performance.

The National Aeronautics and Space Administration (NASA), via its Navigation and Ancillary Information Facility (NAIF) at the Jet Propulsion Laboratory (JPL), manages the SPICE (not an acronym) information system to track space-borne instruments \citep{Acton:1996, Acton:2018}. SPICE libraries~\citep{Annex2020} were used to compute position vectors for IBEX and celestial objects. In addition, a Python script read archival maneuvering logs, describing the inertial pointing vectors of the spacecraft across the mission. We determined when each object (Sun, Earth, or Moon) is in the IBEX field of view (FOV) by comparing its central angular position to a reference value of the detector's FOV angular diameter. The FOV diameter for excluding the Moon and Sun is taken to be 14 degrees based on empirical measurement: if the center of those objects, as a point source, is in the FOV, the data are excluded. The Moon's angular size, about a degree, is small enough that a more detailed calculation is not needed. Although the Sun has an extended corona, the same approximation also proved sufficient, because IBEX is almost never oriented such that the detector faces sunward. For the Earth, a point-source approximation does not suffice, because the magnetosphere is also a source of interference. We instead model the magnetosphere as extending 12 Earth radii from the Earth's apparent location in SPICE, and excluded any time when that includes that region in the FOV. Because light-travel times are short relative to object motion, we do not utilize SPICE's aberration feature for reasons of computational cost. After the position and orientation vectors were ascertained, the AstroPy library~\citep{astropy:2013,astropy:2018,astropy:2022} was used to convert between celestial coordinate systems and units.

Checking whether an object is in the IBEX FOV is the core operation for data selection, yet it occurs in the context of over a decade of observations on a spinning spacecraft moving through Earth orbit. Additional code integrates the FOV check into this context. The process reads IBEX SPICE data and the inertial pointing vectors, and it computes the right ascension and declination coordinates of each pointing vector for the entire mission using AstroPy. The data are validated to avoid duplicated maneuver entries. A set of evenly-spaced sampling times is constructed, typically 48 spins (one IBEX spacecraft rotation), much shorter than the multiple days between most maneuvers. For every sampled time, using the corresponding maneuver's inertial pointing vector, the angular coordinates of the IBEX spin axis and the target (Sun, Earth, or Moon) are computed, along with the angular separation between them. If the target is entirely beyond the FOV of IBEX throughout its rotation, then the data is kept. Alternatively, if the target is inside the FOV, a more detailed check is performed. This check references IBEX spin bins, the 6-degree subdivisions of the sky used to record ENA data. The bins are consistently numbered with respect to the ecliptic north pole; this numbering scheme circumvents difficulty with precise timing on the spacecraft clock. For this check, the IBEX and target angular orientation vectors are converted barycentric mean ecliptic coordinates. The spin bin nearest the target is determined, and all adjacent bins within the IBEX FOV are marked as bad. This approach is slightly over-conservative for targets outside the spin plane.

After a final de-duplication stage, the list of 6-degree bad bins is analyzed to produce a set of good data. The bad bins are parsed, checked for wrapping between bins 0 and 59, and sorted to produce a final list of good bins at constant time intervals that are at least one IBEX FOV away from the target (including the target's magnetosphere, in the case of Earth). This list of good bins for one target is combined with a set of other lists for other targets: together, the Sun-free, Earth-free, and Moon-free lists provide a filter to assert that the times that those bins are free of any contamination by those celestial bodies. This filter simplifies the task of machine learning by removing one of the most significant known sources of noise for the ENA detector.

\section{Table of Features Used in Random Forest}           %
\label{App:RF_Feat}                                               %
\begin{longtable}{||P{.32\linewidth}||P{0.65\linewidth}||}
    \hline
    \textbf{Feature} & \textbf{Description} \\ [0.5ex] 
    \hline\hline
    ESA Sweep & 1 of 6 overlapping energy passbands at which ENAs or background may be observed \\ [1ex]
    \hline 
    Angle Bin & 1 of 60 six-degree bins that inform on the position of IBEX within a given 3-dimensional circular slice of the sky \\ [1ex]
    \hline
    Time & Time associated with a given ESA (true or background) observation \\ [1ex]
    \hline
    Orbit & An approximately 4.5 day period over which IBEX collects data \\ [1ex]
    \hline
    Counts &  The number of ENAs (true or background) observed \\ [1ex]
    \hline
    Counts (Lower) &  Lower threshold for background detection \\ [1ex]
    \hline
    Counts (Upper) &  Upper threshold for background detection \\ [1ex]
    \hline
    Earth Not Visible &  An indicator variable that informs on whether the Earth (or its magnetosphere) is within IBEX's field of view \\ [1ex]
    \hline
    Moon Not Visible &  An indicator variable that informs on whether the Moon is within IBEX's field of view \\ [1ex]
    \hline
    Sun Not Visible &  An indicator variable that informs on whether the Sun is within IBEX's field of view \\ [1ex]
    \hline
    Sum of Counts in angle bin &  Sum of counts within an angle bin, over time (within an orbit, within an ESA) \\ [1ex]
    \hline
    Mean of Counts in angle bin &  Mean of counts within an angle bin, over time (within an orbit, within an ESA) \\ [1ex]
    \hline
    Variance of Counts in angle bin &  Variance of counts within an angle bin, over time (within an orbit, within an ESA) \\ [1ex]
    \hline
    Sum of Counts in Time &  Sum of counts within a time interval, over angle bins (within an orbit, within an ESA) \\ [1ex]
    \hline
    Mean of Counts in Time &  Mean of counts within a time interval, over angle bins (within an orbit, within an ESA) \\ [1ex]
    \hline
    Variance of Counts in Time & Variance of counts within a time interval, over angle bins (within an orbit, within an ESA) \\ [1ex]
    \hline
    Mean Theta Counts Ratio &  ratio of average count observed within a time interval to the minimum average count observed within a time interval for all time intervals (within an orbit, within an ESA) \\ [1ex]
    \hline
    Sum of Counts across ESAs & Sum of counts over ESAs within the same angle bin and time interval (within an orbit) \\ [1ex]
    \hline
    Mean of Counts across ESAs & Mean of counts over ESAs within the same angle bin and time interval (within an orbit) \\ [1ex]
    \hline
    Variance of Counts across ESAs & Variance of counts over ESAs within the same angle bin and time interval (within an orbit) \\ [1ex]
    \hline
    Upper Neighbor's Count  & Number of ESAs (true or background) observed in one angle bin greater than considered observation, but within the same time interval  \\ [1ex]
    \hline
    Lower Neighbor's Count & Number of ESAs (true or background) observed in one angle bin smaller than considered observation, but within the same time interval \\ [1ex]
    \hline
    Right Neighbor's Count & Number of ESAs (true or background) observed in one time interval greater than considered observation, but within the same angle bin \\ [1ex]
    \hline
    Left Neighbor's Count & Number of ESAs (true or background) observed in one time interval smaller than considered observation, but within the same angle bin \\ [1ex]
    \hline
    Upper Left Neighbor's Count &  Number of ESAs (true or background) observed in one time interval greater and one angle bin greater than considered observation \\ [1ex]
    \hline
    Upper Right Neighbor's Count &  Number of ESAs (true or background) observed in one time interval smaller and one angle bin greater than considered observation \\ [1ex]
    \hline
    Lower Right Neighbor's Count &  Number of ESAs (true or background) observed in one time interval greater and one angle bin smaller than considered observation \\ [1ex]
    \hline
    Lower Left Neighbor's Count &  Number of ESAs (true or background) observed in one time interval smaller and one angle bin smaller than considered observation \\ [1ex]
    \hline
    \caption{Features considered in Random Forest}
    \label{Tab:Features}
\end{longtable}

\section{Orbit Plots with Labels and Probabilities: Orbit 471}           %
\begin{figure}[H]
    \centering
    \begin{subfigure}[b]{0.47\textwidth}
        \centering
        \includegraphics[width=\textwidth]{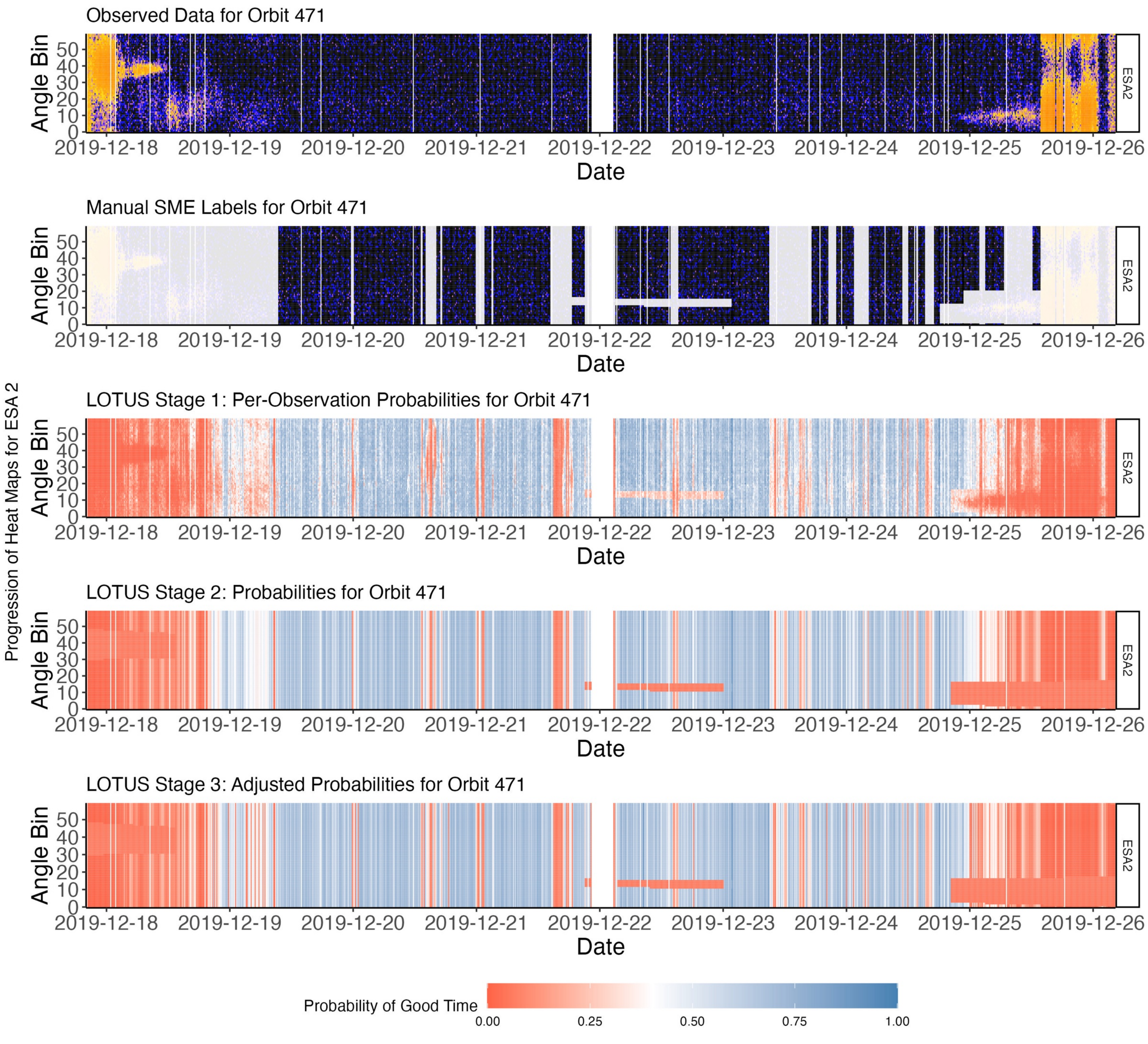}
        \caption{}
        \label{SubFig:471_2_Probs}
    \end{subfigure}
    \begin{subfigure}[b]{0.47\textwidth}
        \centering
        \includegraphics[width=\textwidth]{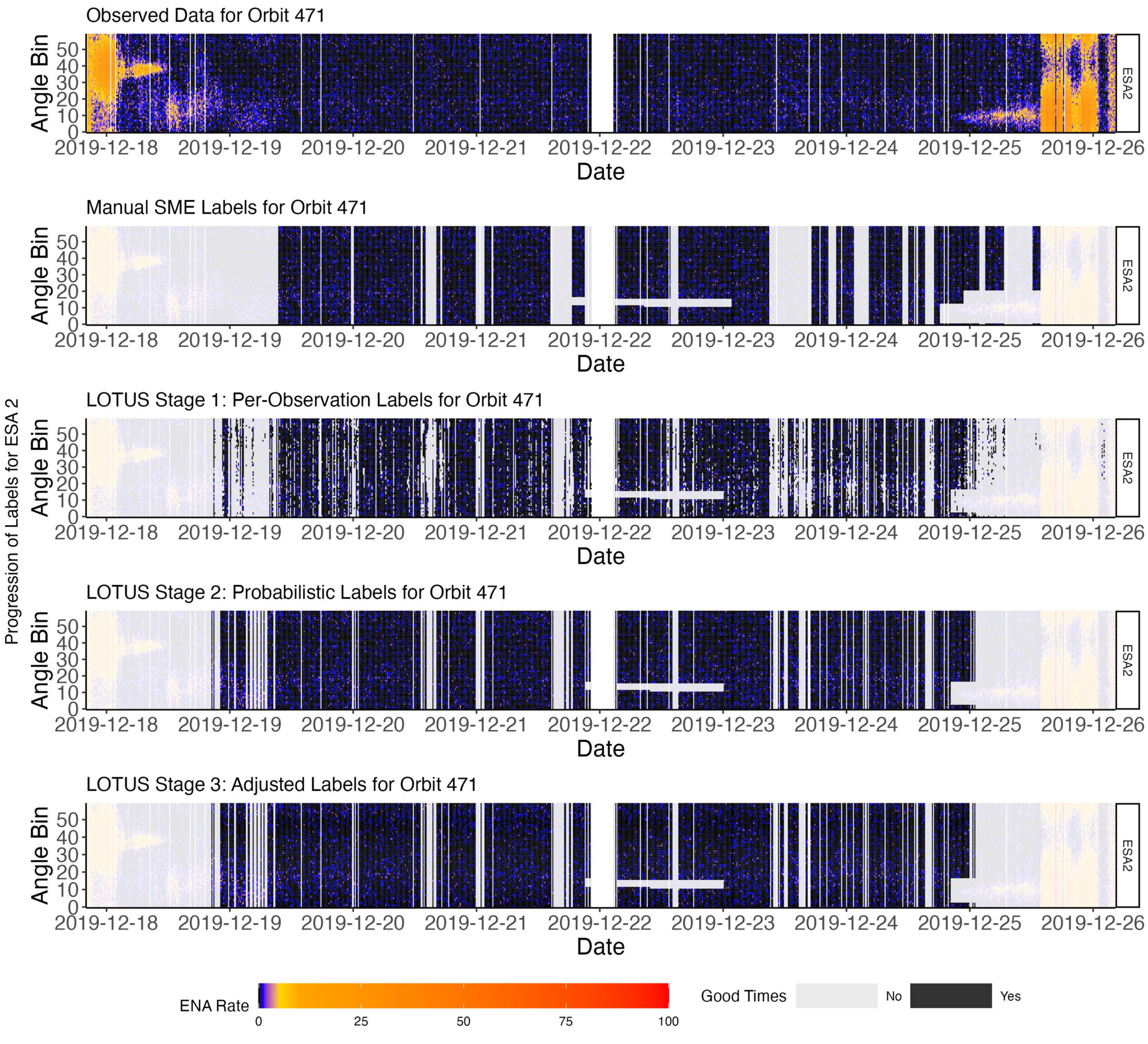}
        \caption{}
        \label{SubFig:471_2_Labs}
    \end{subfigure}
    \caption{Raw data, manual SME labels, and progression of LOTUS Stage 1, 2, and 3 for orbit 471, ESA 2. Row 1 corresponds to the raw data observed by IBEX. Row 2 corresponds to the labels assigned by the SME. Rows 3, 4, and 5 correspond to the outputs of LOTUS Stage 1, 2, and 3, respectively. (a) Probabilistic progression of LOTUS. (b) Label progression of Lotus based on probabilistic output in (a).}
    \label{Fig:471_2_LOTUS}
\end{figure}

\begin{figure}[H]
    \centering
    \begin{subfigure}[b]{0.47\textwidth}
        \centering
        \includegraphics[width=\textwidth]{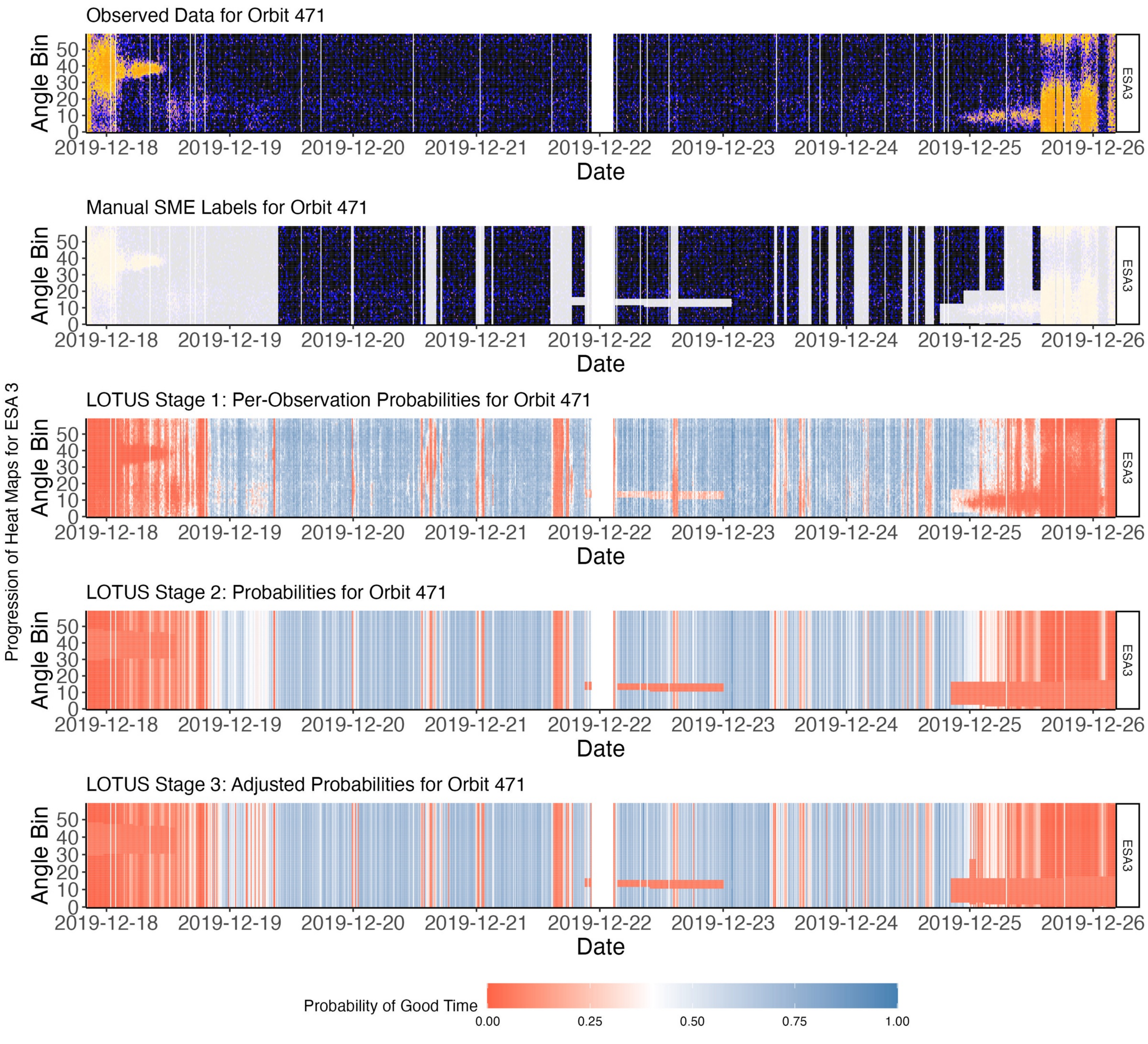}
        \caption{}
        \label{SubFig:471_3_Probs}
    \end{subfigure}
    \begin{subfigure}[b]{0.47\textwidth}
        \centering
        \includegraphics[width=\textwidth]{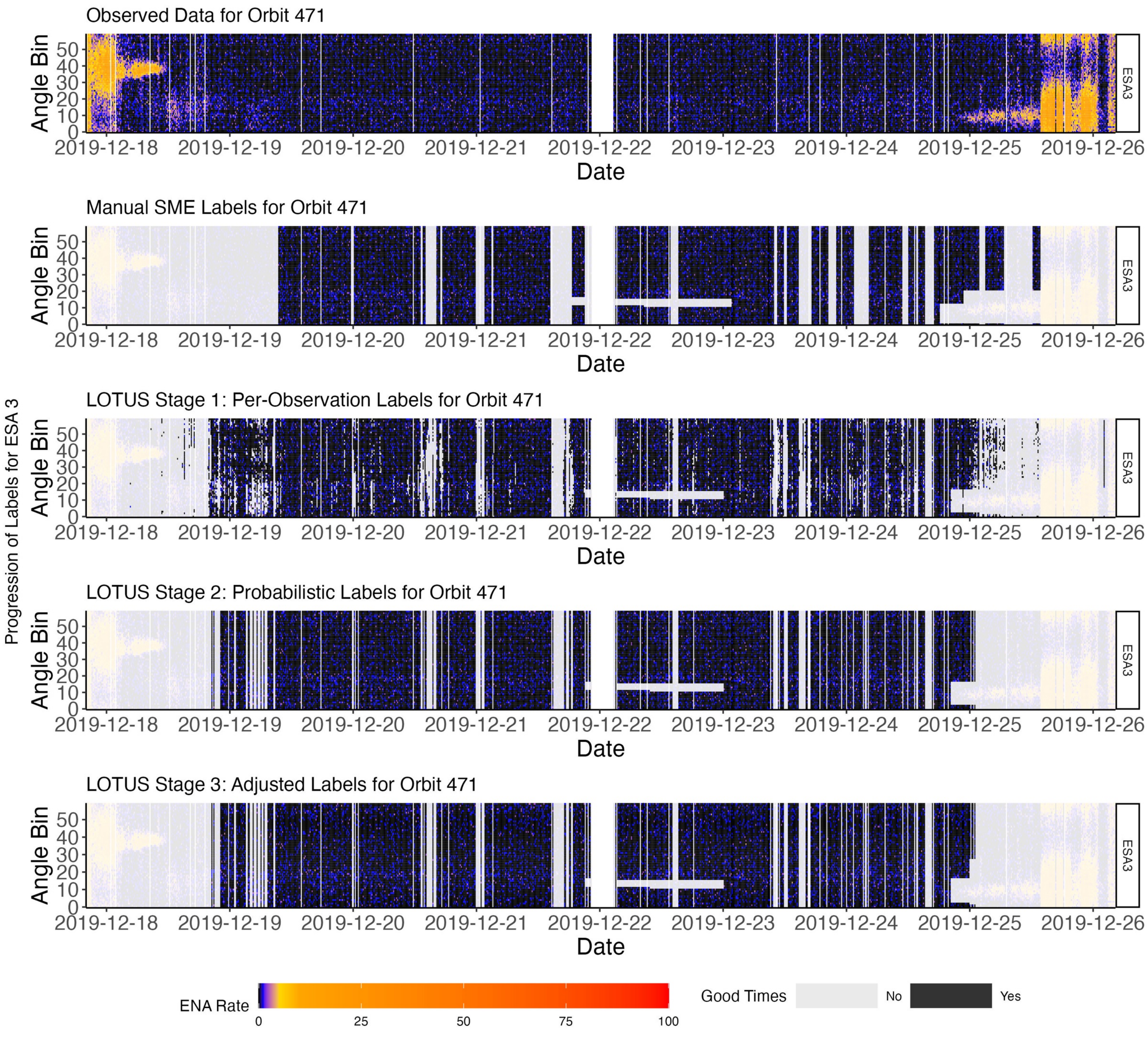}
        \caption{}
        \label{SubFig:471_3_Labs}
    \end{subfigure}
    \caption{Raw data, manual SME labels, and progression of LOTUS Stage 1, 2, and 3 for orbit 471, ESA 3. Row 1 corresponds to the raw data observed by IBEX. Row 2 corresponds to the labels assigned by the SME. Rows 3, 4, and 5 correspond to the outputs of LOTUS Stage 1, 2, and 3, respectively. (a) Probabilistic progression of LOTUS. (b) Label progression of Lotus based on probabilistic output in (a).}
    \label{Fig:471_3_LOTUS}
\end{figure}

\begin{figure}[H]
    \centering
    \begin{subfigure}[b]{0.47\textwidth}
        \centering
        \includegraphics[width=\textwidth]{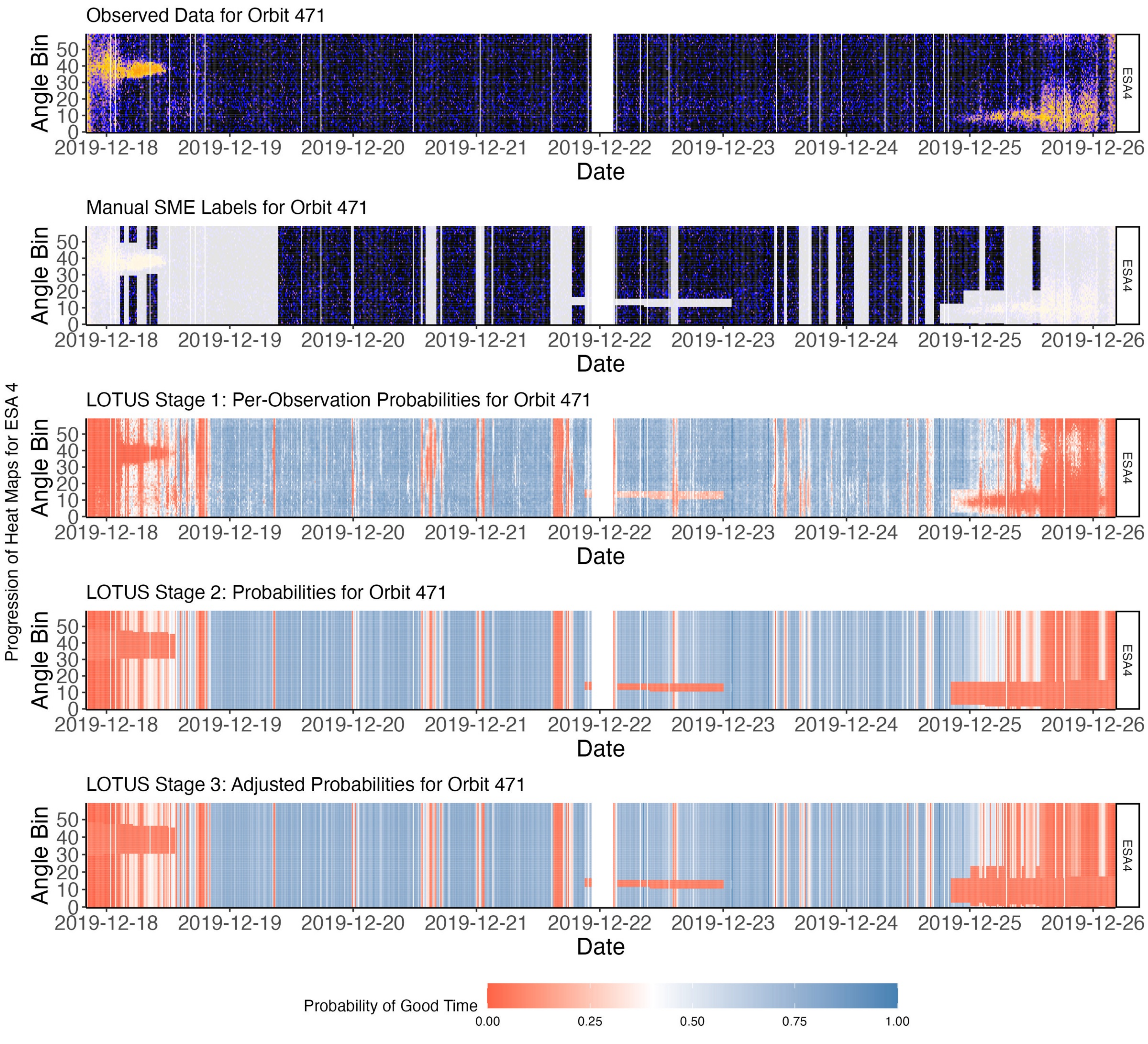}
        \caption{}
        \label{SubFig:471_4_Probs}
    \end{subfigure}
    \begin{subfigure}[b]{0.47\textwidth}
        \centering
        \includegraphics[width=\textwidth]{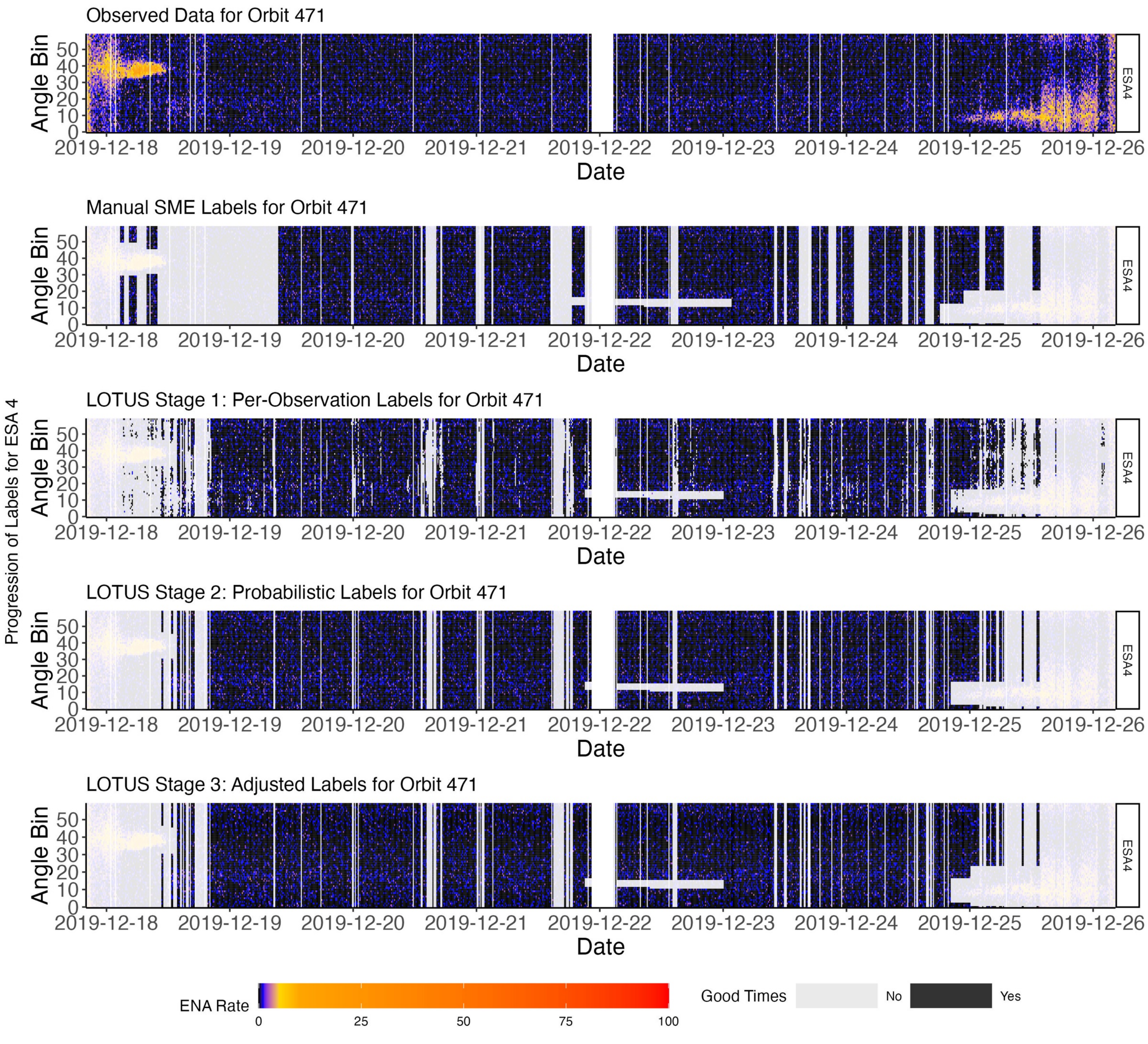}
        \caption{}
        \label{SubFig:471_4_Labs}
    \end{subfigure}
    \caption{Raw data, manual SME labels, and progression of LOTUS Stage 1, 2, and 3 for orbit 471, ESA 4. Row 1 corresponds to the raw data observed by IBEX. Row 2 corresponds to the labels assigned by the SME. Rows 3, 4, and 5 correspond to the outputs of LOTUS Stage 1, 2, and 3, respectively. (a) Probabilistic progression of LOTUS. (b) Label progression of Lotus based on probabilistic output in (a).}
    \label{Fig:471_4_LOTUS}
\end{figure}

\begin{figure}[H]
    \centering
    \begin{subfigure}[b]{0.47\textwidth}
        \centering
        \includegraphics[width=\textwidth]{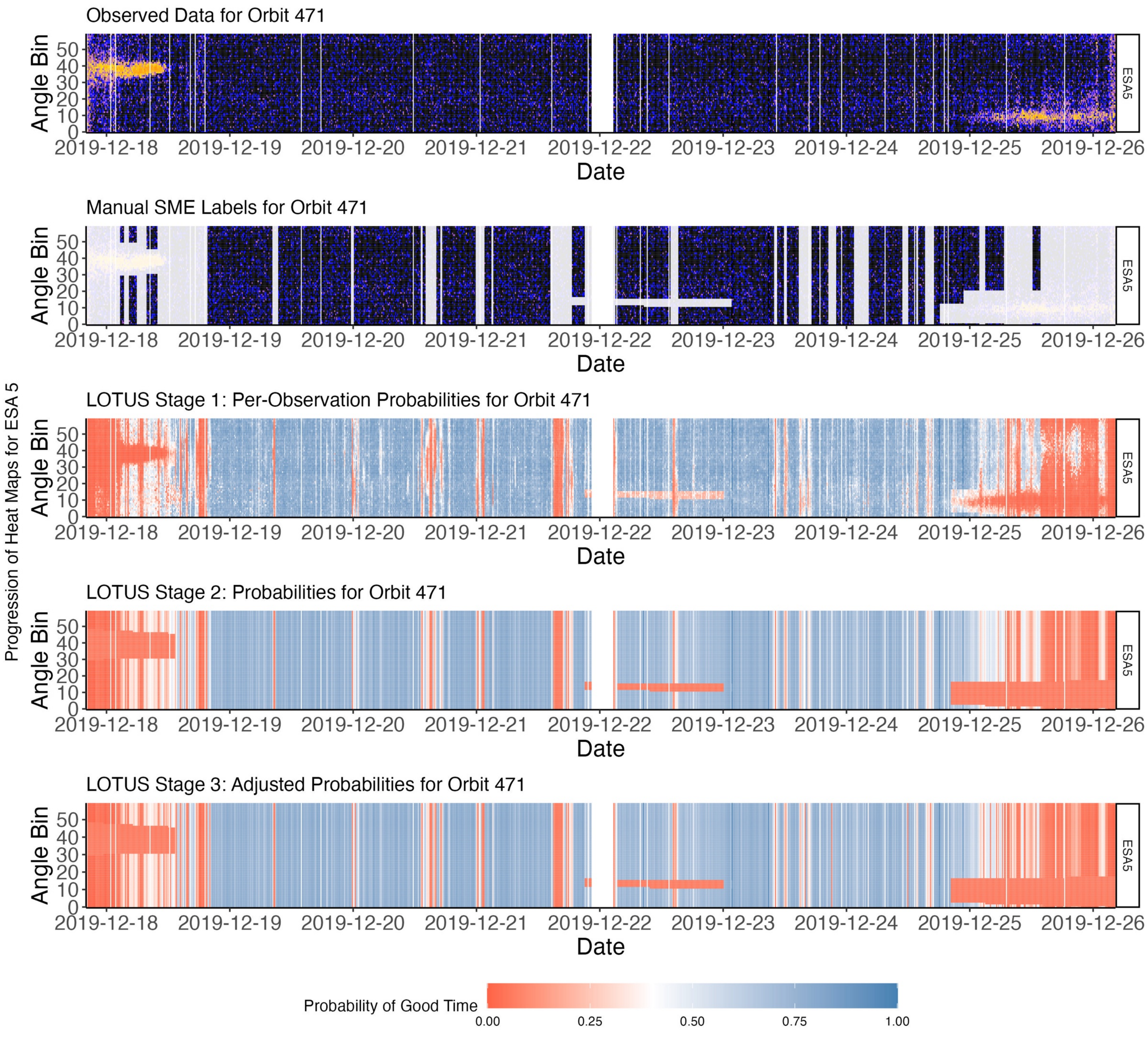}
        \caption{}
        \label{SubFig:471_5_Probs}
    \end{subfigure}
    \begin{subfigure}[b]{0.47\textwidth}
        \centering
        \includegraphics[width=\textwidth]{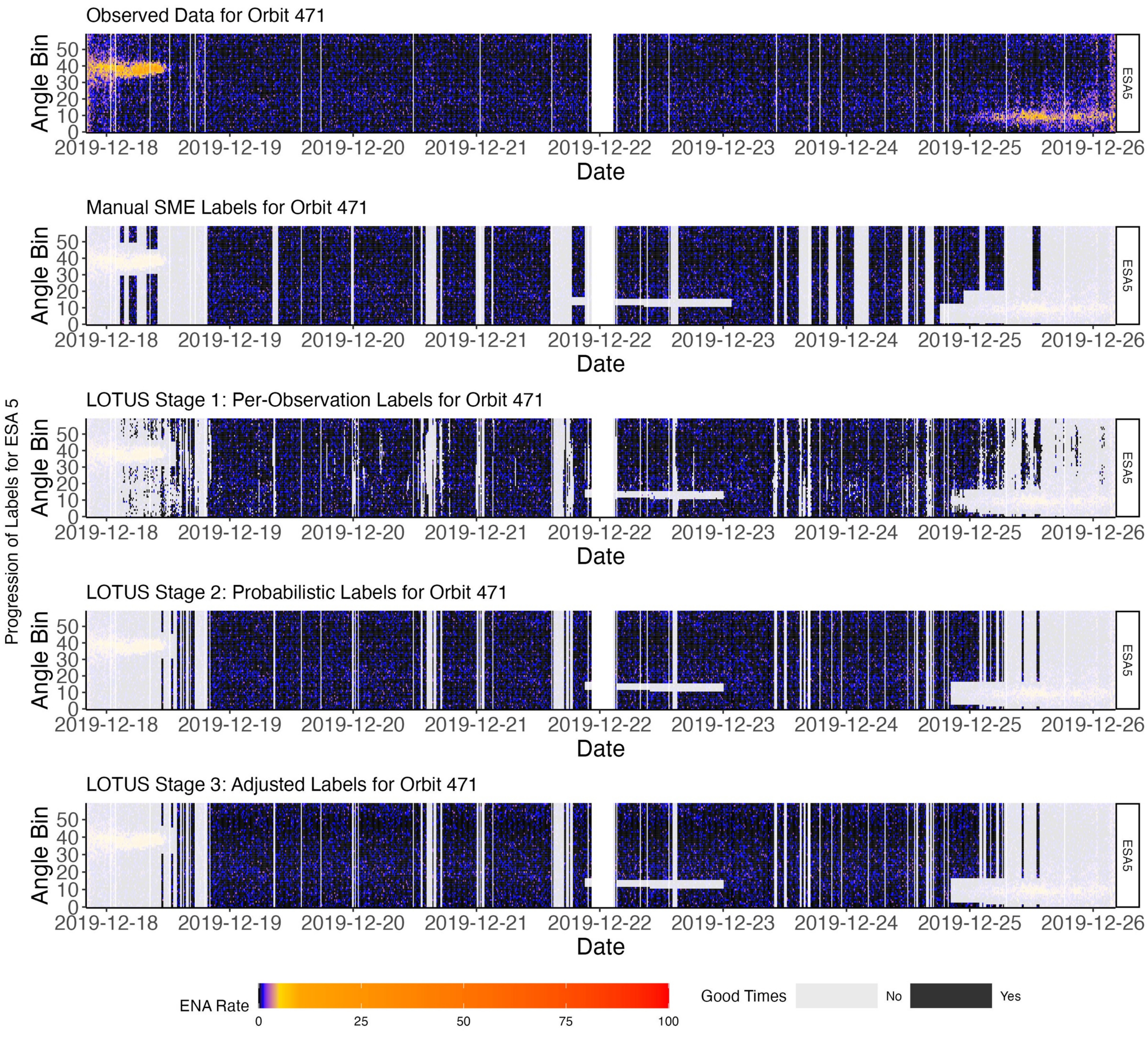}
        \caption{}
        \label{SubFig:471_5_Labs}
    \end{subfigure}
    \caption{Raw data, manual SME labels, and progression of LOTUS Stage 1, 2, and 3 for orbit 471, ESA 5. Row 1 corresponds to the raw data observed by IBEX. Row 2 corresponds to the labels assigned by the SME. Rows 3, 4, and 5 correspond to the outputs of LOTUS Stage 1, 2, and 3, respectively. (a) Probabilistic progression of LOTUS. (b) Label progression of Lotus based on probabilistic output in (a).}
    \label{Fig:471_5_LOTUS}
\end{figure}

\label{App:471_orbitPlots}                                               %
Orbit plots for orbit 471, ESAs 2, 3, 4, and 5 are provided below. Each figure considers the probabilistic output of LOTUS on the left, with the corresponding labels on the right. Figures~\ref{Fig:471_2_LOTUS} through~\ref{Fig:471_5_LOTUS} demonstrate that LOTUS tends to identify the same areas as ``good times'' the SME, with a few additional areas included. As long as the ENA rates correspond, including additional areas is not necessarily bad, and, in fact, produces count rates with smaller statistical uncertaintyl. See Section~\ref{SubSec:ENADat} for more information on ENA rates. 

\end{document}